\documentclass[11p]{elsarticle}
\usepackage{natbib,geometry,fleqn,graphicx}
\usepackage{hyperref,float}
\usepackage{blindtext}
\usepackage{makecell}
\usepackage{caption}
\biboptions{sort&compress}
\usepackage{placeins} % used to allow \floatbarrier
\usepackage{stfloats, cuted, caption}
\usepackage{xcolor}
\makeatletter
\def\ps@pprintTitle{%
	\def\@oddfoot{}%
	\def\@evenfoot{}%
}
\makeatother
\begin{document}
\begin{frontmatter}
\title{Steady inertial flow of a compressible fluid in a spatially periodic channel 
under large pressure drops: a multiscale semi-analytical approach}
\author[sap]{Valentina Biagioni}
\author[vub]{Bram Huygens}
\author[sap]{Giuseppe Procopio}
\author[sap]{Maria Anna Murmura}
\author[vub]{Gert Desmet}
\author[sap]{Stefano Cerbelli\corref{cor1}}
\ead{stefano.cerbelli@uniroma1.it}
\cortext[cor1]{}
\address[sap]{{Dipartimento di Ingegneria Chimica -
Sapienza Università di Roma },
            {Via Eudossiana 18},
            {Roma},
            {00184},
            {},
            {Italy}}
\address[vub]{{Dept. of Chemical Engineering  -
Vrije Universiteit Brussel},
             {Bd de la Plaine 2, 1050 Ixelles},
            {Brussel},
            {1050 Ixelles},
            {},
            {Belgium}}
\begin{abstract}
Spatially-periodic channels are increasingly attracting attention
as an efficient alternative to packed 
columns for a number of analytical  and engineering processes. 
In incompressible flows, the periodic geometry
allows to compute the flow structure by solving the Navier-Stokes (NS)
equations in the minimal periodic cell of the structure, however large the pressure gradient. 
Besides, when gas flow under large pressure drops are dealt with, the velocity
field is not periodic because of the density dependence on pressure.
In this case,  the momentum balance equations
must be solved numerically on the entire channel, thereby requiring
massive computational effort. 
Based on the marked separation of scales between the length of the periodic cell
and the overall channel length characterizing many applications 
we develop a general method for predicting both the large-scale
pressure and velocity profiles, 
and the small-scale flow structure.  
The approach proposed is based on the assumption that the local dimensionless pressure drop
as a function of the Reynolds number, say $g({\rm Re})$, 
can be estimated from the solution of the incompressible
NS equations within the minimal periodic cell of the channel. From the  
knowledge of $g({\rm Re})$,
the pressure profile $P(Z)$ vs 
the large scale axial coordinate $Z$
is derived analytically by quadratures. We show how 
qualitatively different profiles $P(Z)$ can be obtained depending on the
equation of state the gas.
The approach is validated by comparing the predicted profiles with the 
full-scale numerical solution of the 
compressible NS equations in different axially-symmetric periodic 
channel geometries.
\end{abstract}

%\item Research highlight  or obstacles1

%\begin{keyword}
%\end{keyword}

\end{frontmatter}

\section{Introduction}

Microchannels are attracting increasing interest in numerous analytical and engineering fields, including chemical reactors \cite{li2021high, li2024efficient, feng2022residence, chen2024hydrogen}, heat transfer devices \cite{bucak2022heat, rahman2024assessment}, and chromatography \cite{biagioni2022taming}, due to their capability of enhancing heat and mass transfer rates while avoiding the presence of moving mechanical components. An accurate description of these devices requires the solution of mass, heat, and momentum balance equations. With regards to momentum, issues may arise in the description of flow structure due to the significant pressure gradients that often arise. Clearly, this issue is not relevant in the case of incompressible fluids; however, it becomes important when compressibility cannot be neglected. 

The need to accurately describe the flow of compressible fluids has been familiar for some time. 
A consistent body of work has been developed for rarified gas conditions, due to the
important implications as regards momentum, mass and energy transport in ordered and disordered
media whose pores are characterized
by a wide distribution of lengthscales (see, e.g., \cite{Farkya2025,Kosyanchuk202290,Zhu2017,Patronis2014532}). 
In this context, the gaseous nature of the fluid mainly impacts upon the structure
of the velocity profile at the solid surfaces confining the flow, thereby determining
a variety of fluid dynamic regimes, from continuum, to
slip, transitional, and, ultimately, free-molecular flow, as the Knudsen number increases from order
$10^{-2}$ to unity and beyond.
 
A comparatively smaller number of studies 
focused on relatively high pressure conditions to unveil the effect of the coupling between
density and pressure on the flow structure in empty channels.
For instance, Proud'homme et al. 
\cite{prud1986laminar} developed a two-dimensional solution for the velocity and 
pressure profiles in steady, laminar, isothermal flow of an ideal gas in  long 
tubes as a double perturbation expansion in the radius to length ratio of the tube, 
$\beta$, and the relative pressure drop, $\epsilon$. Their results 
validate the Hagen-Poiseuille assumption of local fluid incompressibility for the 
obtainement of the pressure profile. Similar approaches were employed by Venerus 
\cite{venerus2006laminar, venerus2010compressible}. An experimental study of the 
problem was carried out by Celata et al. 
\cite{celata2007experimental,celata2009friction}, 
who determined the friction factor for a gas flow in a microchannel 
and found an excellent agreement with the Hagen-Poiseuille correlation.
The direct numerical solution through Finite Element Method (FEM) approach 
of gas flow in single empty microchannels \cite{guo2008numerical}
and 
in intercrossing \cite{bejhed2006numerical} microchannel networks 
has been also reported. Because of the multiscale nature of the problem, 
however, in these studies  the ratio of the characteristic dimension of the cross-section to
the channel length was limited to $10^{-2}$.
Another aspect that must be considered is the range of Reynolds number for which the flow of a 
compressible fluid is laminar. This problem was dealt with by Novopashim et al. 
\cite{novopashin2016laminar}, who studied the effect of the value of the second Virial 
coefficient on the critical Reynolds number for which there is a transition from 
laminar to turbulent flow of a gas in a cylindrical channel and found that appreciable 
differences in the values of the critical Reynolds numbers exist at high pressures. 
Vocale et al. \cite{vocale2022numerical}, on the other hand, carried out a numerical 
study on the effect of fluid compressibility on the friction factor in microchannels 
having a square cross-section. They found that compressibility leads to an increase 
in the average friction factor, and that the value of the Reynolds number at which 
these effects become significant decreases in microchannels with smaller cross-section.

As regards the case of periodic channels, 
the importance of making available a theoretical and computationally affordable framework
for predicting the flow structure
can hardly be overestimated. Periodic channel geometries (e.g.~obtained using
Triply Periodic Minimal Surfaces as solid fillers embedded in the channel
\cite{su2025experimental,cheng2021investigations,jespers2017chip,bragin2024flow}) 
exploited in microturbine
cooling, or
micropillar array columns used for gas chromatography
\cite{meziani2022evaluation,liu2021behavior} provide
just two out of many examples that could greatly
benefit from an accurate description of the small-scale
and large-scale structure of the compressible flow. In these examples,
the detailed knowledge of the small- and large-scale structure of the flow
could be exploited for enforcing simplifying assumptions
in the mass and energy transport equations,
thus allowing to interpret experimental results of existing prototypes, on the one hand,
and/or to define optimal shapes for enhancing the heat transfer coefficient in micro heat exchangers,
or lowering 
the axial dispersion coefficient in gas chromatography.

More generally,
the periodic geometry template could serve as a an idealized model of random packing
\cite{hermann2014geometric,thabet2018development,lester2013chaotic},
thus  enlarging the domain of application of the approach proposed to classical
chemical engineering equipment such as adsorption columns and fixed bed gas-solid
or catalytic reactors.
In this perspective, the assessment of the fluid dynamics features of the gas
flow under large pressure drops in periodic media constitutes 
the preliminary yet fundamental 
piece of information allowing to define a new class of 
problems in 
Brenner's theory of macrotransport processes in periodic
media \cite{brenner1993macrotransport}, where the effective
transport coefficients resulting from the intertwined
action of molecular diffusion and convection at the 
scale of single cell of the periodic structure 
are not constant throughout the medium even in the 
isothermal case because of
their depencence on the large-scale pressure profile.

The above observations highlight both the 
importance of understanding and accurately describing 
the flow of compressible fluids in periodic channels 
and the potential opportunities of employing 
these devices for heat and mass transfer operations.
Specifically, it is foreseeable that 
the detailed knowledge of the large- and small-scale structure
of the velocity field in the compressible case will make it possible to devise geometries
and operating conditions tailored to control mixing, dispersion
and heat transport in gas flows. This would allow to trace over 
the results
obtained for microchannels processing liquid solutions,
where the strictly spatially periodic structure of the flow
has been exploited for investigating the relationship 
between flow structure and residence time
distribution\cite{poumaere2022residence,raynal2014distribution}, 
and enhancing the performance of micromixers,  
microreactors\cite{lester2018simultaneous,gorodetskyi2014analysis,habchi2013chaotic,jilisen2013three} and 
microfluidics-assisted separation columns for liquid chromatography
\cite{gelin2021reducing,de2012realization}. 
 
To the best of these authors' knowledge, 
however, a computationally affordable method for 
describing the flow of compressible fluids 
in such channels has never been proposed. In the present work, we attempt to fill this gap
 by developing a method for predicting the large-scale pressure and velocity profiles 
downstream a periodic channel, as well as the small-scale flow structure,  
taking advantage of the simplifying assumptions made possible by the significant separation 
of scales between the single periodic cell and the overall channel length that characterize typical applications. 
We show that an accurate description of the flow is made possible by a multiscale 
approach, where the gas velocity is split into a large-scale piecewise-constant 
factor and a strictly periodic incompressible component. The approach proposed
hinges on the observations that, notwithstanding the considerable
large variations of density, velocity and pressure that may arise (depending on the geometry) downstream the
channel axis,
mass conservation implies that at steady-state conditions the cross-sectional average of the product
between the  density and the axial velocity component is constant along the channel axis. On the assumption
that the viscosity variations can be neglected, this allows to define
a Reynolds number characterizing the entire flow, and to effectively decouple the dependence of the
variables on the the large- and the small-scale coordinates.
 
The article is organized as follows. In Section 2, we describe the problem being 
dealt with and illustrate the channel geometries considered as case studies 
along with the definition of the coordinate systems adopted. In Section 3 the 
theoretical approach adopted is discussed. The results are presented in Section 4 and discussed 
in Section 5, together with a perspective towards possible generalizations of the approach to
non-isothermal flows and/or random media.

\section{Statement of the problem}
\begin{figure}[H]
        \centering
        \includegraphics[width=0.65\linewidth]{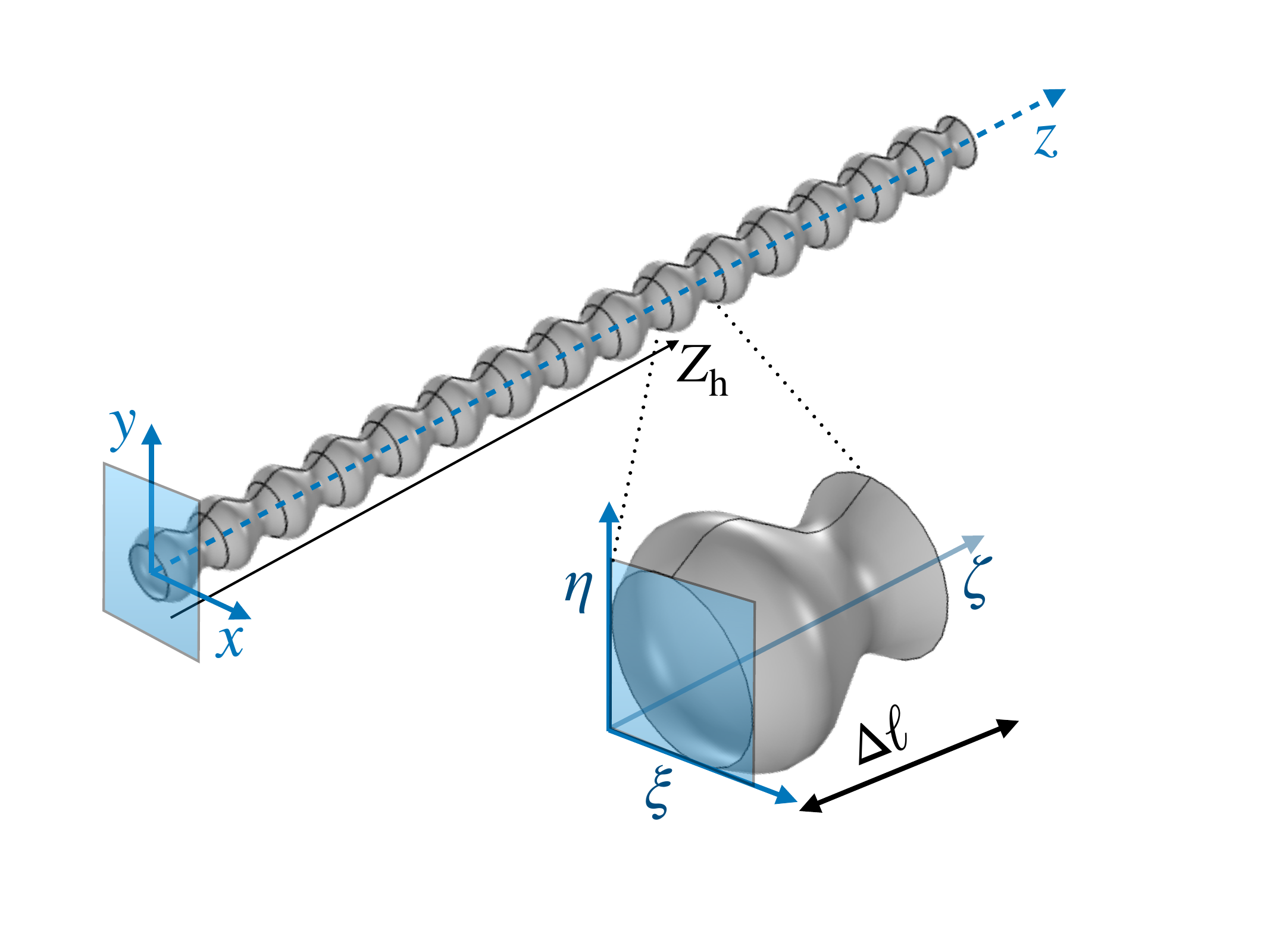}
        \caption{Schematic representation of the periodic channel geometry
and of the global and local coordinate systems, $x$, $y$, $z$ and ${\xi},{\eta},{\zeta}$, respectively.}
        \label{figzero}
\end{figure}

We consider
the pressure-driven, compressible flow of a gas through an axially periodic 
channel. Figure \ref{figzero} depicts an example of periodic channels, together
with the global and local (or cell) coordinate systems.

In what follows, the spatial period of the geometry, say ${\Delta}{\ell}$, is assumed
to be orders of magnitude smaller than the overall length, $L$, of the channel, 
i.e.~${\Delta}{\ell}/L <<1$. 
Also, the flow is assumed Newtonian, steady, non-turbulent, and isothermal (see Section 5.1 for more details on this assumption).
In the incompressible case,
the velocity field
can be obtained through the solution of Navier-Stokes equations on the minimal periodic
cell of the structure, thereby assuming that, beyond a short spatial transient, 
the flow is periodic over the same length
${\Delta}{\ell}$ defining the geometry of the system. 
In the case of compressible fluids, however,
the velocity field cannot be expected in general to be periodic. This is because the existence
of a pressure drop along the channel axis causes the simultaneous variation
of the fluid density and (by mass conservation) of the average
axial velocity.
The flow structure must therefore be obtained by solving
 the entire channel structure of the
compressible Navier-Stokes Boundary Value Problem,
\begin{equation}
{\rho}  \mathbf{v} \cdot {\nabla}   \mathbf{v}  
={\mu}{\nabla}^2 \mathbf{v}+\frac{\mu}{3} \, 
{\nabla} \big ( {\nabla} \cdot \mathbf{v} \big ) - {\nabla}p \mbox{;} 
\qquad {\nabla} \cdot \big ({\rho}  \mathbf{v} \big )=0
\label{nseq}
\end{equation}
equipped with no-slip Boundary Conditions (BCs) onto all the solid surfaces 
confining the flow, and
with assigned pressure values at the inlet and outlet cross-sections of the channel.
In Eq.~(\ref{nseq}), ${\rho}(x,y,z)$,  $\mathbf{v}(x,y,z)$, $p(x,y,z)$ represent
the local density, velocity and pressure of the gas. 
The constant density constraint of the incompressible case is
here replaced by the Equation Of State (EOS), $f({\rho},p, T)=0$, that characterizes
the volumetric behavior of the gas. In the case of widely separated scales, where
${\Delta}{\ell}/L$ is of order $10^{-3}$  or below, the direct numerical approach to 
Eq.~(\ref{nseq}) may result troublesome. This is especially true  
in 
inherently three-dimensional flows in the inertial regime because
of the possible onset of 
thin boundary layers, which, even at the scale of the single periodic module, require
a large number of degrees of freedom to be captured. Thus, for channels embedding a large
number of cells (e.g.~order $10^3 \div 10^5$), the computational effort 
becomes massive, making it impractical to obtain
accurate numerical solutions of the flow when geometries of practical interest  are to be dealt with.
\subsection{Case studies}
\begin{figure}[H]
        \centering
        \includegraphics[width=0.75\linewidth]{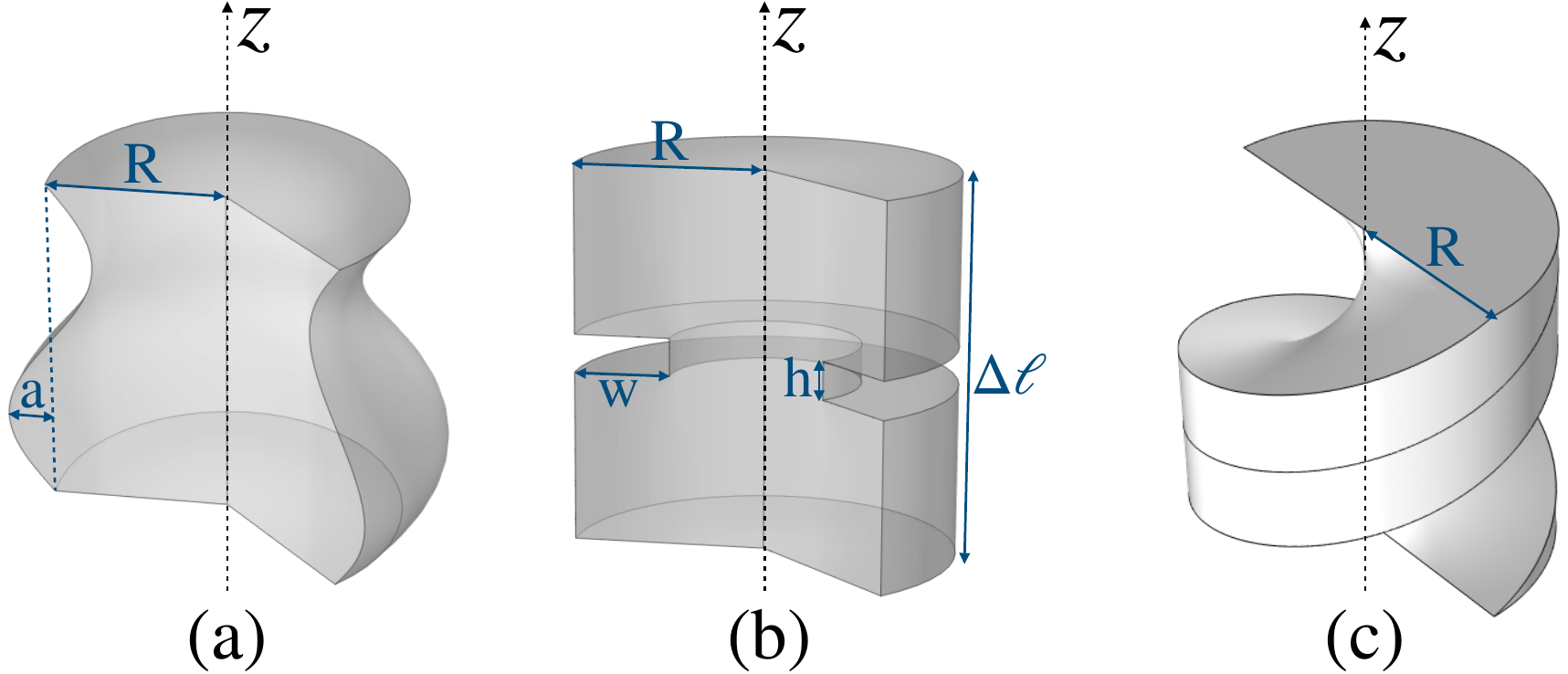}
        \caption{Elementary cells of the periodic channel geometries 
considered as case studies.} 
        \label{figone}
\end{figure}

As case studies, the three geometries depicted in Fig.~\ref{figone}
are next considered. 
The channel shapes depicted in panels (a) and (b) 
of the figure are axial-symmetric and will be used to compare and
contrast the predictions stemming from the approach proposed in this article
with the full solution of the Navier-Stokes problem expressed by 
Eq.~(\ref{nseq}) for the compressible isothermal flow of an ideal gas.
In all of the cases represented, the aspect ratio of the periodic cell has been chosen
so that ${\Delta}{\ell}=2R$.
The wall of the sinusoidal channel of Fig.~\ref{figone}-(a)
is specified by the equation
\begin{equation}
\sqrt{\tilde{x}^2+\tilde{y}^2}=\tilde{R}+\tilde{a}  \sin ( 2{\pi}\tilde{z})
\label{sinchanneq}
\end{equation}
where $\tilde{x}=x/{\Delta}{\ell}$, $\tilde{y}=y/{\Delta}{\ell}$ and
$\tilde{z}=z/{\Delta}{\ell}$ are nondimensional coordinates scaled
to the length of the periodic cell, and where
the parameters $\tilde{R}=R/{\Delta}{\ell}$ and $\tilde{a}=a/{\Delta}{\ell}$ represent the 
dimensionless average
channel radius and the amplitude  of the sinusoidal oscillation, respectively. \textcolor{black}{In the case investigated, the dimensionless amplitude of the sinusoidal profile has been fixed equal to $\tilde{a} = 0.125$.}
As regards the baffled channel (panel b), the depth, $w$, and thickness, $h$, of the wall indentation
have been chosen so that $w/R=1/2$ and $h/{\Delta}{\ell}=1/10$.
The shape in Fig.~\ref{figone}-(c) is obtained by inserting a helicoidal
baffle in a cylindrical channel (for better visualization,
only one of the two halves of the
structure is depicted in the figure). The helicoidal surface is specified 
by the parametric equations
\begin{equation}
\left \{
\begin{array}{lll}
\tilde{x} & = & u \, \cos(2{\pi}v)\\
\tilde{y} & = & u \, \sin(2{\pi}v)\\
\tilde{z} & = & v 
\end{array}
\right .
\label{helixeq}
\end{equation}
where $0 \leq u \leq R/{\Delta}{\ell}$ ($R$ being the radius of the cylinder)
and $0 \leq v \leq L/{\Delta}{\ell}$.
The helicoidal channel of Fig.~\ref{figone}-c will be taken as a representative 
example of a three-dimensional periodic flow where the 
inertial term of momentum balance equation shows a strong influence 
of the streamline geometry even at relatively low values of the Reynolds 
number.

\section{Theoretical approach and  
analytical prediction of pressure and velocity large-scale profiles 
}
\subsection{Decomposition of the field variables in small- and large-scale components}
Tracing over classical approaches of homogenization theory \cite{majda1999simplified,brenner1993macrotransport},
we represent the axial coordinate $z$ as $z=Z_h+\tilde{z}$, where
$Z_h=[z/{\Delta}{\ell}] \, {\Delta}{\ell}$ 
(``$[u]$'' being the integer part of $u$) identifies
the distance of the $h$-th cell from the channel inlet (see Fig.~\ref{figzero}). 
The variable 
$0 \leq \tilde{z} \leq {\Delta}{\ell}$ is a local coordinate spanning the axial
extent of the cell, i.e.~$\tilde{z}=(z-Z_h)$.
In the approach next proposed the fluid is supposed locally incompressible,
thus implying that ${\rho}(x,y,z)$ is represented by a piecewise-constant function, 
\begin{equation}
{\rho}(x,y,z)=R(Z_h)
\label{rhospliteq}
\end{equation}
undergoing finite jumps when crossing the boundary between two contiguous periodic cells.
By enforcing the above structure for the the density field, in the Appendix we show that an approximation
to the full-scale solution of the compressible Navier-Stokes problem in the periodic geometry for both velocity
and pressure can be expressed in terms of the product of large-scale piecewise-constant functions times a small-scale
(dimensionless) field that captures the fluctuations of the field variables at the characteristic length
of the cell size.

As regards the velocity, this is represented as 
\begin{equation}
\mathbf{v}(x,y,z)=V(Z_h) \, \mathbf{v}_c(x,y,\tilde{z})
\label{vspliteq}
\end{equation}
and
where $V(Z_h)$[m/s] represents the cell-averaged value  of the large-scale component
of the axial velocity,
\begin{equation}
V(Z_h)=\int_{V^{(h)}_c} \mathbf{v} \cdot \mathbf{n}_z \; d{x}
 \, d{y} \, d{z}  \big /  \int_{V^{(h)}_c} \; d{x} \,  d{y}  \, d{z}
 \label{eq5}
 \end{equation}
and where $\mathbf{v}_{c}(x,y,\tilde{z})$ is a dimensionless vector-valued
function periodic in the interval $0 \leq \tilde{z} \leq {\Delta}{\ell}$.
In Eq.~(\ref{eq5}), $V^{(h)}_c$ denotes the volume of the generic
$h$-th cell downstream the channel and $\mathbf{n}_z$ is a unit vector aligned with the z-axis.

The pressure, $p(x,y,z)$, is also 
decomposed into a large-scale component, say $P(Z_h)=R(Z_h) V^2(Z_h)$, which is assumed to scale as the
density of the kinetic energy, times a small-scale dimensionless pressure field, as
\begin{equation}
p(x,y,z)=R(Z_h) \, V^2(Z_h) \left ( 
p_{c}({x},{y},\tilde{z})-{\Delta}p^{\prime} \, \tilde{z} \right )
\label{pspliteq}
\end{equation}
where $p_{c}({x},{y},\tilde{z})$ is periodic in $(0,{\Delta}{\ell})$, and
where ${\Delta}p^{\prime}$ is a dimensionless  
pressure drop across the $h$-th cell.
Note that by Eq.~(\ref{pspliteq}), the product  
$ R(Z_h) \, V^2(Z_h) \, {\Delta}p^{\prime}$ yields the large-scale
pressure drop across the $h$-th cell, which must be necessarily different from 
zero if an average axial velocity component is to be present.

As shown in the Appendix, the small-scale fields $\mathbf{v}_{c}(x,y,\tilde{z})$
and $p_{c}({x},{y},\tilde{z})-{\Delta}p^{\prime} \, \tilde{z}$ are the
($h$-independent) solutions of the dimensionless Navier-Stokes incompressible
problem 
\begin{equation}
\, \mathbf{v}_c \cdot {\nabla}_c  \mathbf{v}_c =
\frac{1}{\rm Re} \bigg ( {\nabla}^2_c\mathbf{v}_c+ 
({1}/3) {\nabla}_c \big ( {\nabla}_c \cdot \mathbf{v}_c \big )       \bigg )
-{\nabla}_c p^{\prime} \mbox{;} \qquad \nabla \cdot  \mathbf{v}_c =0
\label{dimnseq}
\end{equation}
defined over the generic periodic cell volume, henceforth denoted by $V_c$,
equipped with no-slip boundary conditions onto all of the solid surfaces
confining the flow, and with periodic boundary conditions for the
velocity at the opposite cross-section delimiting the unit cell.
Under the assumption that
the dynamic viscosity of the gas can be regarded as a constant through the entire channel,
the Reynolds number
\begin{equation}
{\rm Re}= \frac{R(Z_h) \, V(Z_h) \, {\Delta}{\ell}}{{\mu}}
\label{Reeq}
\end{equation}
entering Eq.~(\ref{dimnseq}) is independent of the cell location $h$ since mass conservation
at steady state conditions implies that (see the Appendix)
\begin{equation}
R(Z_k)\, V(Z_k)=R(Z_l)\, V(Z_l)
\label{massconseq}
\end{equation}
for any couple of integers $0 \leq k,l \leq N_{\rm tot}$, where
$N_{\rm tot}$ is the total number of cells covering the channel length.
In the representation of the field variables described above,  
the average mass flux, say $\dot{m}$, through any cross-section 
of the generic $h$-th cell is thus given by
\begin{equation}
\dot{m}=R(Z_h)\, V(Z_h)=\mbox{const.}
\label{avmassfluxeq}
\end{equation}
from which ${\rm Re}=\dot{m}{\Delta}{\ell}/{\mu}$.
One notes that the dimensioless pressure drop ${\Delta}p^{\prime}$ and the 
Reynolds number, ${\rm Re}$ cannot be fixed independently of each other in that
the average axial velocity,$V(Z_h)$ is a function  of the the dimensional pressure drop
over a generic $h$ cell, $R(Z_h) \, V^2(Z_h) \, {\Delta}p^{\prime}$.
This implies that the dimensionless pressure drop ${\Delta}p^{\prime}$ across the 
cell is uniquely determined by the Reynolds number and the 
geometry of the periodic module, and
thus well-defined relationship between the cell pressure drop and Reynolds number of the type
\begin{equation}
{\Delta}p^{\prime}=g({\rm Re})
\label{gReeq}
\end{equation}
must exist, where the function $g({\rm Re})$, which is proportional to the cell
friction factor, only depends on the geometry of the periodic cell. 
Equation (\ref{gReeq}) is next used to derive the large-scale profiles of pressure,
velocity and density.
\subsection{Analytical prediction of the large-scale pressure profile}
Consistently with the assumption of widely separated scales,
when addressing the pressure, velocity and density variations along the channel,
the large scale coordinate $Z_h$ can be regarded as a continuous variable, i.e.~$Z_h=Z$. Likewise,
the large-scale pressure drop across the generic $h$-th cell is interpreted as the local gradient of the 
large-scale pressure at the local $Z$ coordinate, i.e.
\begin{equation}
\frac{dP}{dZ} \simeq
\left ( \frac{{\Delta}P}{{\Delta}{\ell}}  \right )_{Z_h}  
=- \frac{R(Z) \, V^2(Z) \, {\Delta}p^{\prime}}{{\Delta}{\ell}} 
= - \frac{R(Z) \, V^2(Z) \, g({\rm Re})}{{\Delta}{\ell}} 
\label{pgradeq}
\end{equation}
Taking into account that $\dot{m}=R(Z_h) \, V(Z_h)=R(Z) \, V(Z)={\rm Re} \, {\mu}/{\Delta}{\ell}$,
 Eq.~(\ref{pgradeq}) can be
re-written as
\begin{equation}
\frac{dP}{dZ}=-  \frac{\dot{m}^2 \, g({\rm Re})}{{\Delta}{\ell} \, R(Z)}=
 - \frac{{{\mu}^2}\; {\rm Re}^2 g({\rm Re})}{{\Delta}{\ell}^3 R(Z)} 
\label{pgradeq1}
\end{equation}
Recalling that ${\rm Re}$ is constant throughout the
entire channel, Eq.~(\ref{pgradeq1}) can be integrated 
by separation of variables if a relationship between the large-scale pressure and density is enforced
and if the dependence of ${\mu}$ on the pressure $P$ is known.
In what follows, we assume that the dynamic viscosity of the fluid is independent of the 
pressure.
The relationship $R=F(P)$ at constant temperature
can be obtained by assuming that the large-scale state variables $P(Z)$ and $R(Z)$ satisfy the
equation of state $f({\rho}^{\prime},p,T)$ of the gas (where
${\rho}^{\prime}$ represents the molar density of the gas), i.e.~$f(R(Z)/M_w,P(Z),T)=0$,
$M_w$ being the average molecular weight of the gas.

For instance, if $f({\rho},p,T)=0$ is given as a virial expansion in the form
\begin{equation}
\frac{p}{{\rho}^{\prime} \mathcal{R} T}= 1+ B^{\prime}(T) \, p+ C^{\prime}(T) \, p^2 + \cdots
\label{virialeq}
\end{equation}
where ${\rho}^{\prime}$ is the molar density of the gas
and $\mathcal{R}$ the universal gas constant, one obtains
\begin{equation}
\frac{R(Z)}{M_w}= \frac{1}{\mathcal{R}T} \frac{P(Z)}{1+B^{\prime}(T) \, P(Z)+ C^{\prime}(T) \, P^2(Z) + \cdots}
\label{lsvirialeq}
\end{equation}
which, substituted in Eq.~(\ref{pgradeq1}) allows to derive the $Z$ dependence of the large-scale
pressure by separation of variables and quadratures.
In the case of an ideal  gas ($B^{\prime}=C^{\prime}=\cdots=0$), Eq.~(\ref{pgradeq1}) yields  
\begin{equation}
\frac{P^2(0)}{2}-\frac{P^2(Z)}{2}=
\frac{\dot{m}^2  g({\rm Re}) \, \mathcal{R} T}{M_W} \, Z
=\frac{{{\mu}^2}\; {\rm Re}^2 g({\rm Re})}{{\Delta}{\ell}^3} \, Z
\label{pprofeq}
\end{equation}
If the entrance and outlet pressure are assigned,
Eq.~(\ref{pprofeq}) can be used to compute the ${\rm Re}$ number,
and thus the mass flow rate through the channel. Conversely, at fixed mass flowrate and  outlet pressure,
the same equation can be used to determine
the inlet pressure necessary to sustain the given mass flowrate.
One notes that Eq.~(\ref{pprofeq})
constitutes the generalization to an arbitrary spatially-periodic channel
of the well-known Hagens-Poiseuille
law for an ideal isothermal gas flowing through an empty straight capillary 
\cite{landau1987fluid}. In the remainder of this article, we
show how the inclusion of a much larger class of channel geometries
may give rise to steady-state gas flows where the inertial term
determines a Re-dependent structure of the streamline geometry. 
This phenomenon finds no counterpart in the Poiseuille case,
where the local incompressibility assumption implies that the flow
is in the Stokes regime until the critical Reynolds value marking
the onset of the first hydrodynamic instability
is reached.  

\section{Results}
All of the results shown below have been obtained by solving the periodic cell 
and the full-scale Navier Stokes problem by the Finite Element Solver Comsol Multiphysics 6.2,
using either tailored regular quadrilateral meshes (axial-symmetric geometries "(a)" and "(b)" of Fig.~\ref{figone}),
or an unstructured tetrahedral mesh (3d helicoidal geometry, panel (c) of the same figure). For the axial-symmetric geometries,
the number, $N$, of degrees 
of freedom was fixed based on the largest affordable discretization of the full-scale problem, which resulted of order $N \simeq 
10^6$. 
The comparison between the multiscale approach-based predicted profiles and the solution of the Navier-Stokes problem 
in the large  was carried out by using identical discretization when solving Eq.~(\ref{nseq}) and Eq.~(\ref{dimnseq}).
 
\subsection{Local pressure drop}
\begin{figure}[H]
        \centering
        \includegraphics[width=0.75\linewidth]{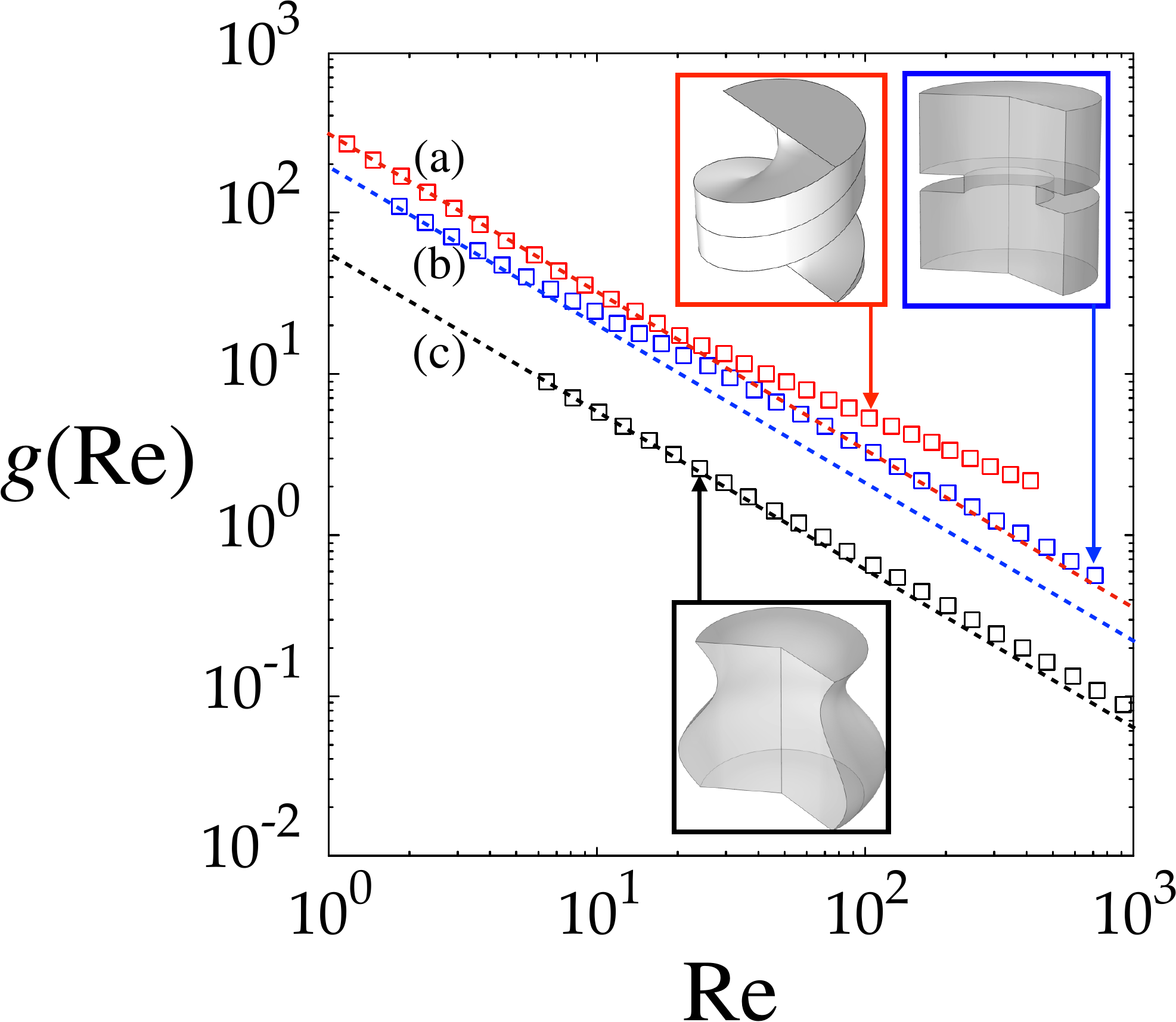}
\caption{Dimensionless pressure drop, ${\Delta}p^{\prime}=g({\rm Re})$ vs ${\rm Re}$ for the three
case studies of Fig.~\ref{figone}.
Black and blue symbols (c) and (b) refer to the sinusoidal and
baffled axial-symmetric geometries, respectively. Red symbols (a) depict ${\Delta}p^{\prime}$ for the 3d helicoidal
geometry. Dashed lines represent the scaling $\mbox{const.}/{\rm Re}$. 
Both the axial-symmetric baffled geometry and the
helicoidal channel depart significantly from the $1/{\rm Re}$ scaling in the range of Reynolds number considered.} 
\label{figthree}
\end{figure}

The implementation of the multiscale approach described above begins with
the solution of the periodic cell problem in the interval of interest of 
values of the Reynolds number, from which the dimensionless pressure drop
$g({\rm Re})$ can be obtained. From a practical point of view, the problem
consists of fixing arbitrarily ${\Delta}{\ell}$, the fluid density and viscosity, and
computing the
periodic flow within the unit cell for
a given interval of values of the pressure drop
${\Delta}p \in (0, {\Delta}p_{\rm max})$.
For a given ${\Delta}p$,
the solution of the incompressible problem determines  the average axial velocity, from which 
the ${\rm Re}$ value associated with ${\Delta}p$ can be computed.
Therefore, the function $g({\rm Re})$ is constructed by simply sweeping the
values of ${\Delta}p$ in the given interval and by computing ${\rm Re}$ and ${\Delta}p^{\prime}$
{\em ex-post}
from the average velocity obtained from the solution and from the fixed values of density and viscosity. 
Figure \ref{figthree} (symbols) shows the results of this computation for the three geometries
depicted in Fig.~\ref{figone}. The dashed lines in the same figure depict the scaling 
$g({\rm Re})=\mbox{const}/{\rm Re}$
pertaining to the case of a Stokes flow, where the 
impact of the nonlinear term in the Navier-Stokes equations is immaterial
and the streamline geometry is independent of Reynolds.
The three cases considered show distinctive responses depending on the specific geometry, 
all characterized by positive  
deviations from the ${\rm Re}^{-1}$
behavior. The magnitude of the deviation increases orderly 
when moving from the sinusoidal channel, to the baffled and the helicoidal
geometry.
To unveil the origin of  the large deviations from the Stokes scaling
exhibited by the helicoidal channel (red symbols in the figure) let us
next consider the dependence of the periodic component, $\mathbf{v}_c$,  of the flow on  
${\rm Re}$ for this specific geometry.
\begin{figure}[H]
        \centering
        \includegraphics[width=0.90\linewidth]{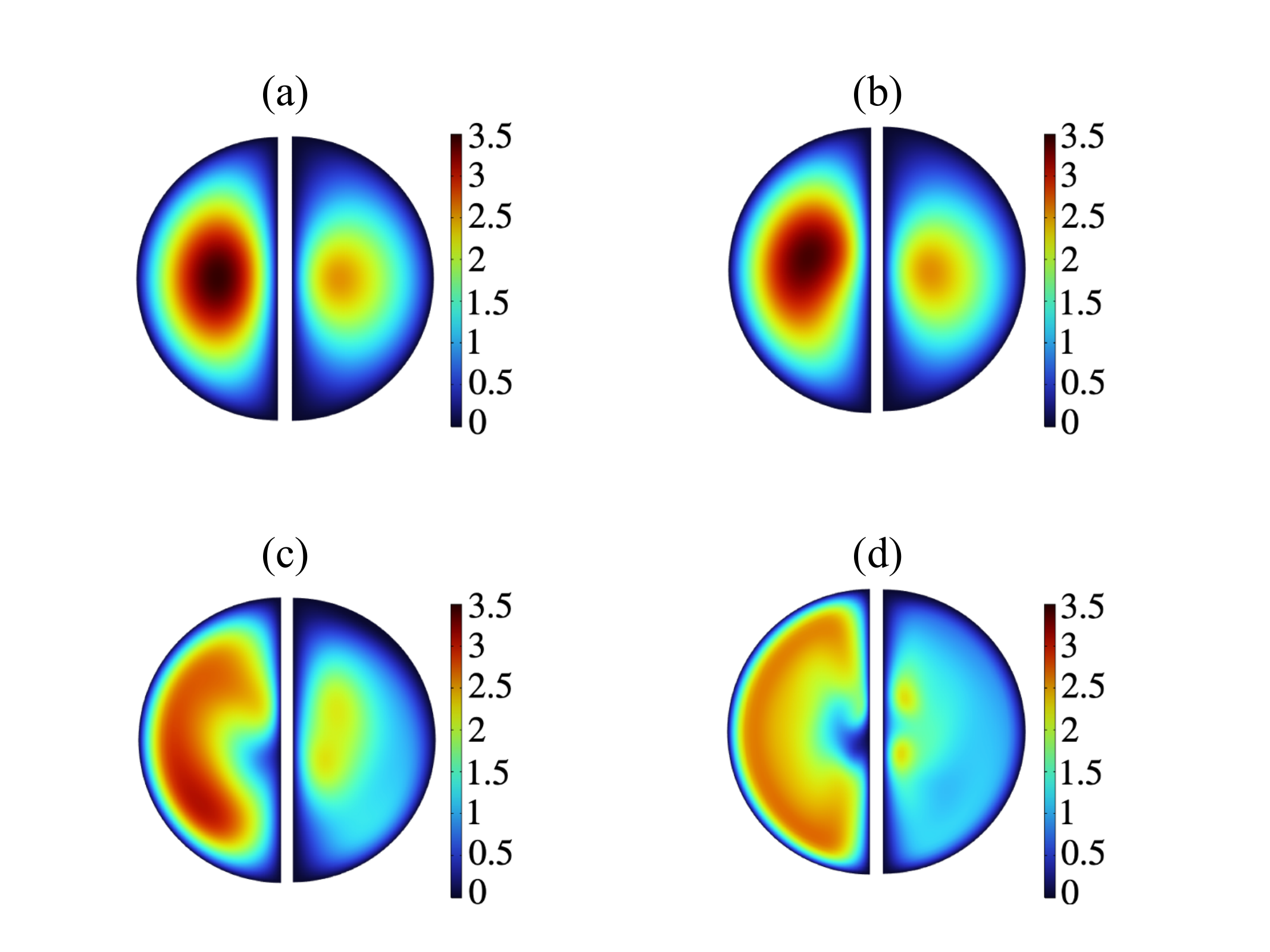}
\caption{Transversal (left semicircle) vs axial (right semicircle) velocity magnitude
for the helicoidal channel (depicted in Figure 2(c)) at increasing values of ${\rm Re}$. Velocity values have been made dimensionless
with respect to the cross-sectional average axial velocity. Panels (a), (b), (c), and (d) depict
the flow structure at ${\rm Re}= \, 1; \; 10; \; 100; \; 450$, respectively.}
\label{figfour} 
\end{figure}
Figure \ref{figfour} depicts the magnitude of the cross-sectional
velocity $v_{c,{\perp}}=\sqrt{v_{c,x}^2+v_{c,y}^2}$ (left semicircles)
vs the axial velocity component, $v_{c,z}$ (right semicircles) at increasing
values of the Reynolds number. As can be gathered from the data, the spatial distribution of the 
cross-sectional and axial velocity magnitude changes both qualitatively and quantitatively in such
a way that the ratio $v_{c,{\perp}}/v_{c,z}$ is not preserved when ${\rm Re}$ varies.
Because the flow streamlines are everywhere tangent to the local velocity vector,
the geometry of the flow as a whole becomes ${\rm Re}$-dependent. 
 
\subsection{Large-scale profiles}
\begin{figure}[H]
        \centering
        \includegraphics[width=0.8\linewidth]{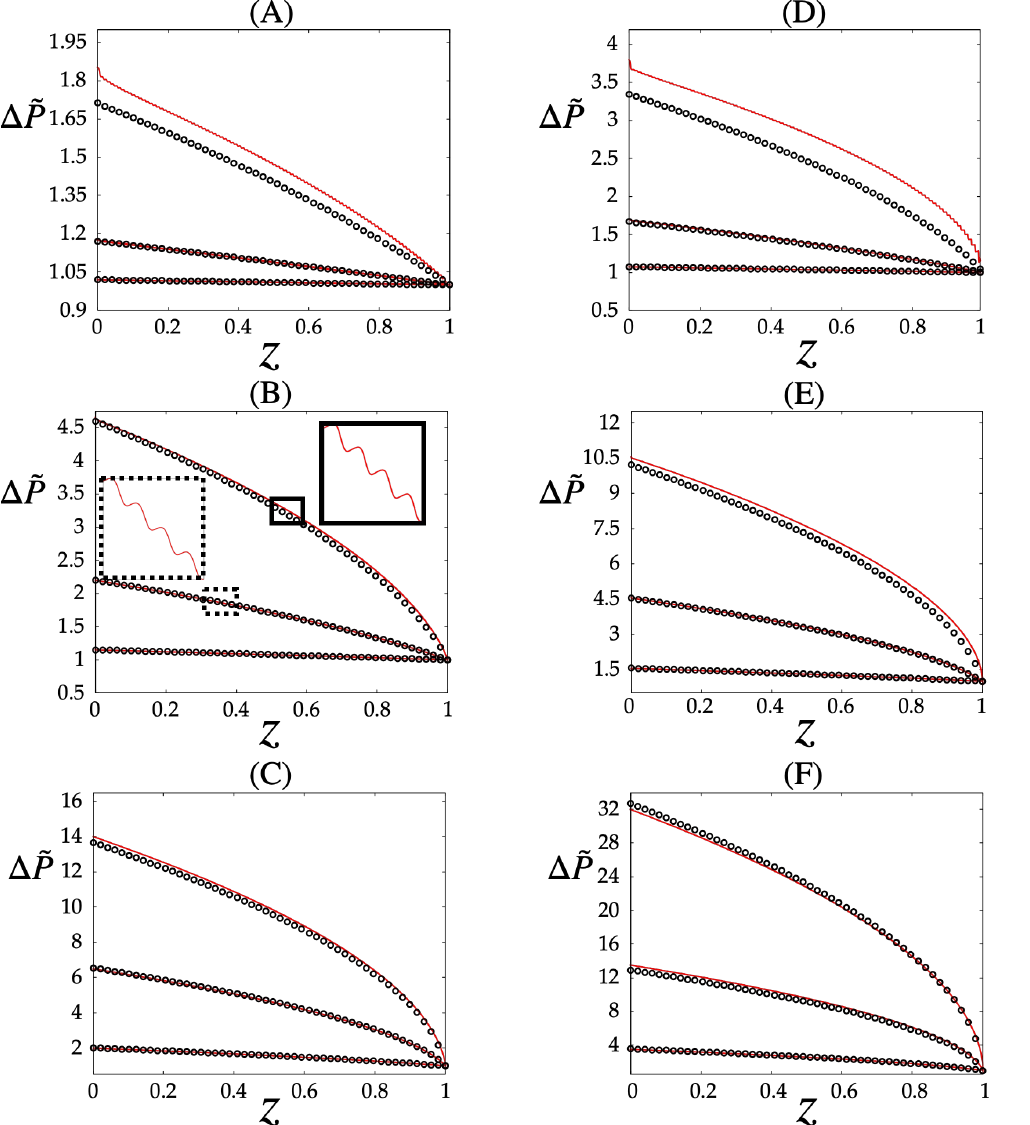}
\caption{Comparison between the large-scale pressure as predicted by the multiscale approach (symbols) and the pressure
profile at the channel axis computed by numerically solving the isothermal full-scale Navier-Stokes problem
on the entire channel structure (continuous lines). The  fluid is assumed to obey the ideal gas law. 
The arrows indicate increasing values or the Reynolds number,
$Re=10,100, 450$. Left column: sinusoidal channel. Right column: baffled channel. Panels (a)-(d), (b)-(e) and (c)-(f)
depict different levels of the scale separation parameter ${\lambda}={\Delta}{\ell}/L$.
(a)-(d): 
${\lambda}=10^{-2}$; (b)-(e): ${\lambda}=10^{-3}$; (c)-(f): ${\lambda}=10^{-4}$. The insets of Panel (b) show the zoomed-in
pressure profile computed at the channel axis as obtained by the full-scale solution of the compressible Navier-Stokes problem
of Eq.~(\ref{nseq}).} 
\label{figfive} 
\end{figure}

\begin{figure}[h]
        \centering
        \includegraphics[width=0.65\linewidth]{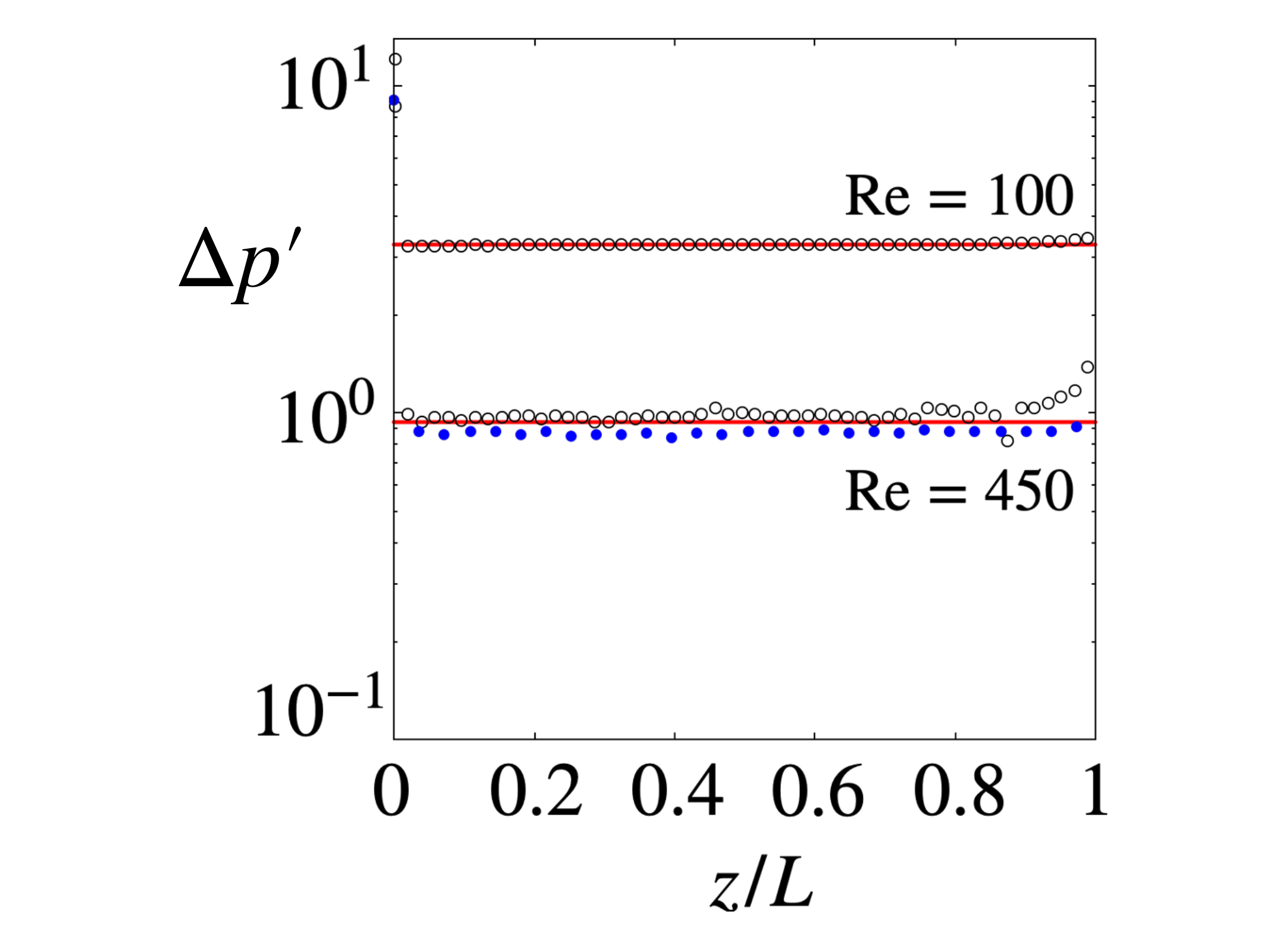}
       \caption{Comparison between the nondimensional pressure drop across a cell
based on the local incompressibility
assumption, $g({\rm Re})$, and the cell pressure drop as computed by the full solution of the
compressible Navier-Stokes equation for the channel geometry of panel (b) of Fig.~\ref{figone}.
Continuous (red) line: $g({\rm Re})$; empty circles: full solution at ${\lambda}=2R/L=10^{-3}$;
Solid (blue) circles: full solution at ${\lambda}=2R/L=10^{-4}$.}
\label{figfive_bis}
\end{figure}

As discussed in Section 3, in the case of isothermal flow the 
knowledge of the local dimensionless pressure drop $g({\rm Re})$
and the enforcement of an equation of state between the large-scale density and pressure, $R(Z)$ and $P(Z)$, respectively,
allows to obtain
the profile by separation of variables and integration of Eq.~(\ref{pgradeq1}).
Next, the prediction of this multiscale approach is compared and contrasted to direct the full solution
of the Navier-Stokes problem in the large expressed by Eq.~(\ref{nseq}), obtained through the Finite-Element software Comsol
Multiphysics (version 6.2). Owing to the large number of degrees of freedom required for solving the latter problem,
only the axially-symmetric structures depicted in panel (a) and (b) of Fig.~\ref{figone}, 
which can be addressed in a two-dimensional computational framework, have been considered.
In this comparison, the volumetric behavior of the compressible fluid is assumed to obey the ideal gas law, thus
yielding the pressure dependence on the large scale axial coordinate expressed by Eq.~(\ref{pprofeq}).
Figure \ref{figfive} shows the dimensionless pressure $P(Z)/P_{\rm out}$ for the sinusoidal channel (left column) and for the
the baffled geometry (right column). Panels (a)-(d), (b)-(e) and (c)-(f) refer to values of the parameter
${\lambda}={\Delta}{\ell}/L$ ranging from $10^{-2}$ to $10^{-4}$, corresponding to channels embedding  $10^2$ to $10^4$ periodic units.
Several observations can be drawn from the data depicted:
\begin{enumerate}
\item The deviation from the predicted and the full-scale solution depends on the separation of scale
parameter ${\lambda}$ as well as on the Reynolds number.
\item At fixed $\rm Re$ value, the departure from a linear pressure scaling (constant large-scale pressure drop)
      depends on the specific cell geometry, as can be observed by comparing corresponding curves between left and right columns.
\item At ${\lambda} \leq 10^{-3}$ the multiscale approach provides an excellent estimate of the full-scale solution at all values of the
Reynolds number considered, relative errors being contained within few percentage points.
For instance, at ${\lambda}=10^{-4}$ the relative error is order 1\%   even in conditions where the relative
pressure variation across the channel is of order 30.
the multiscale approach can either overestimate or underestimate the local pressure value.
\end{enumerate}
A better insight about the validity of the local incompressibility assumption can be gained
by comparing the value of the function $g({\rm Re})$ 
with the local value of the scaled pressure drop ${\Delta}p^{\prime}$,
as computed from the full solution of the compressible Navier-Stokes equations.
Figure \ref{figfive_bis}
shows the result of this comparison for the case of geometry (b) of Fig.~\ref{figone}
As can be observed, beyond minor fluctuations associated with the numerical discretization,
the local pressure drop values are quantitatively
consistent with those computed assuming locally incompressible flow throughout most
of the channel axis. A consistent deviation can be observed  in the 
region near the channel exit, where the local friction factor based on the incompressible assumption
is underpredicted with respect to that computed from the full compressible Navier-Stokes problem.

Once the large-scale pressure profile has been obtained form the multiscale approach, the density $R(Z)$ can be immediately computed
by the equation of state of the fluid, and the large-scale velocity $V(Z)$ can thus be obtained by mass conservation (not discussed in the interest of brevity). 
\subsection{Local flow structure}
\begin{figure}[h]
        \centering
        \includegraphics[width=0.55\linewidth]{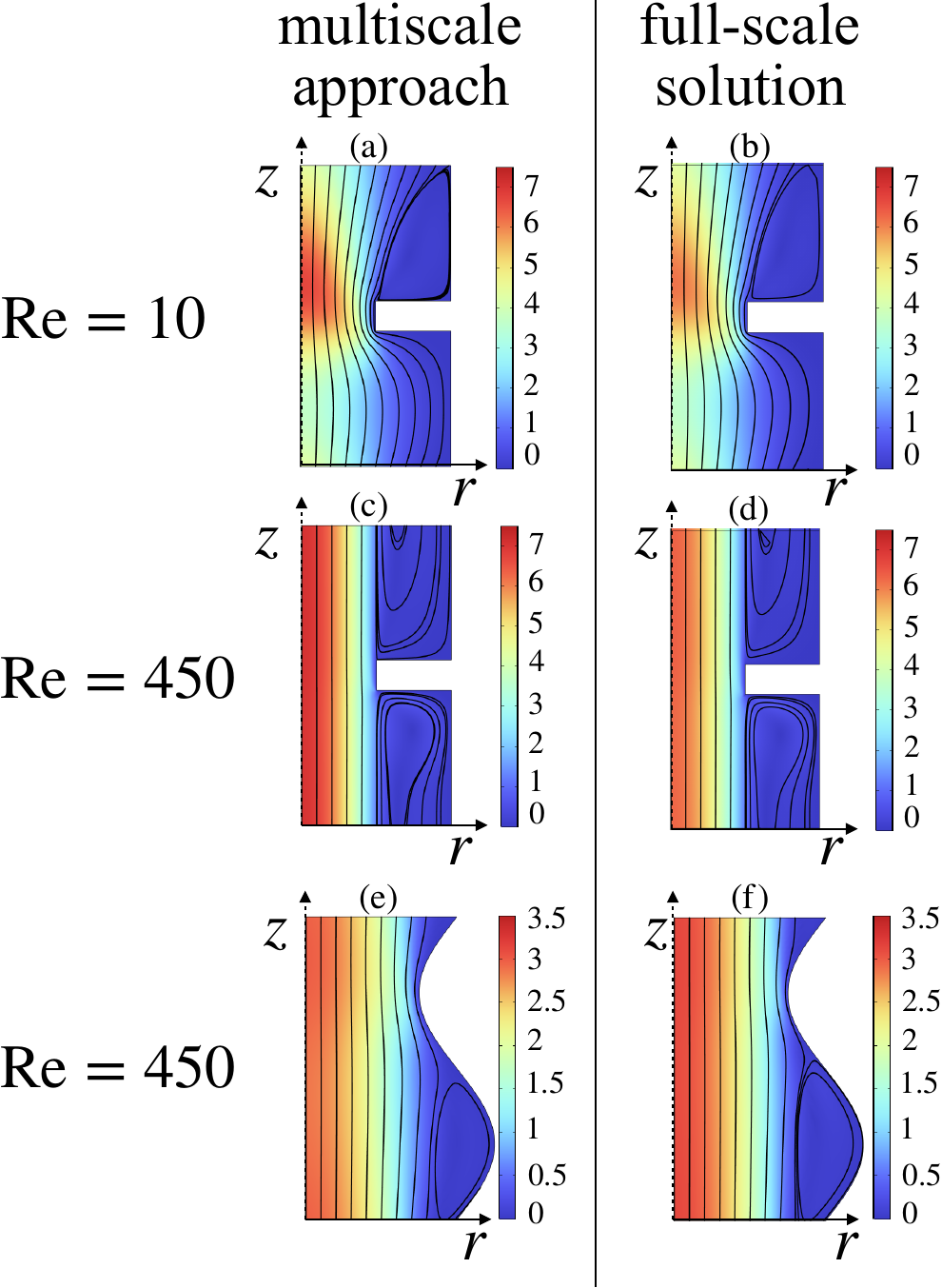}
\caption{Comparison between the small-scale structure of the 
flow as computed by the multiscale approach and the
full-scale solution of the Navier-Stokes compressible flow. The contour colors depict the intensity of the velocity
magnitude scaled to its cell-averaged value. The continuous line represent selected flow streamlines.}
\label{figsix}
\end{figure}

This much established for the large-scale profiles, we next investigate to what extent
the multiscale approach can provide useful information on the small-scale structure
of the flow. This issue is of fundamental importance in applications, in that
the local flow structure controls both the mass and heat transport coefficients, and the very presence of the
periodic geometry, possibly embedding internal obstacles and/or baffles, is often meant 
to obtain increased surface-to-volume ratio as well as convection-enhanced transversal transport.
Figure \ref{figsix} shows the side-by-side comparison of the fine structure of the velocity field
within a periodic cell located halfway through the channel at different ${\rm Re}$ values
as computed by the multiscale approach (left column) and the full-scale solution (right column) of the Navier-Stokes
equations for a value ${\lambda}=10^{-3}$ of the separation of scales parameter in both the sinusoidal and the baffled geometry.
As can be observed, the fine structure of the flow is well represented by the multiscale approach
both as regards the velocity intensity and the flow geometry, 
even  at values of ${\rm Re}$ at which the
non-linear term in the Navier-Stokes equation causes the streamline structure to depart  significantly
from that pertaining to Stokes (creeping) flow conditions.
\section{Discussion}

\subsection{Relevance of isothermal flows and extension to non-isothermal conditions}
\begin{figure}[H]
        \centering
        \includegraphics[width=0.6\linewidth]{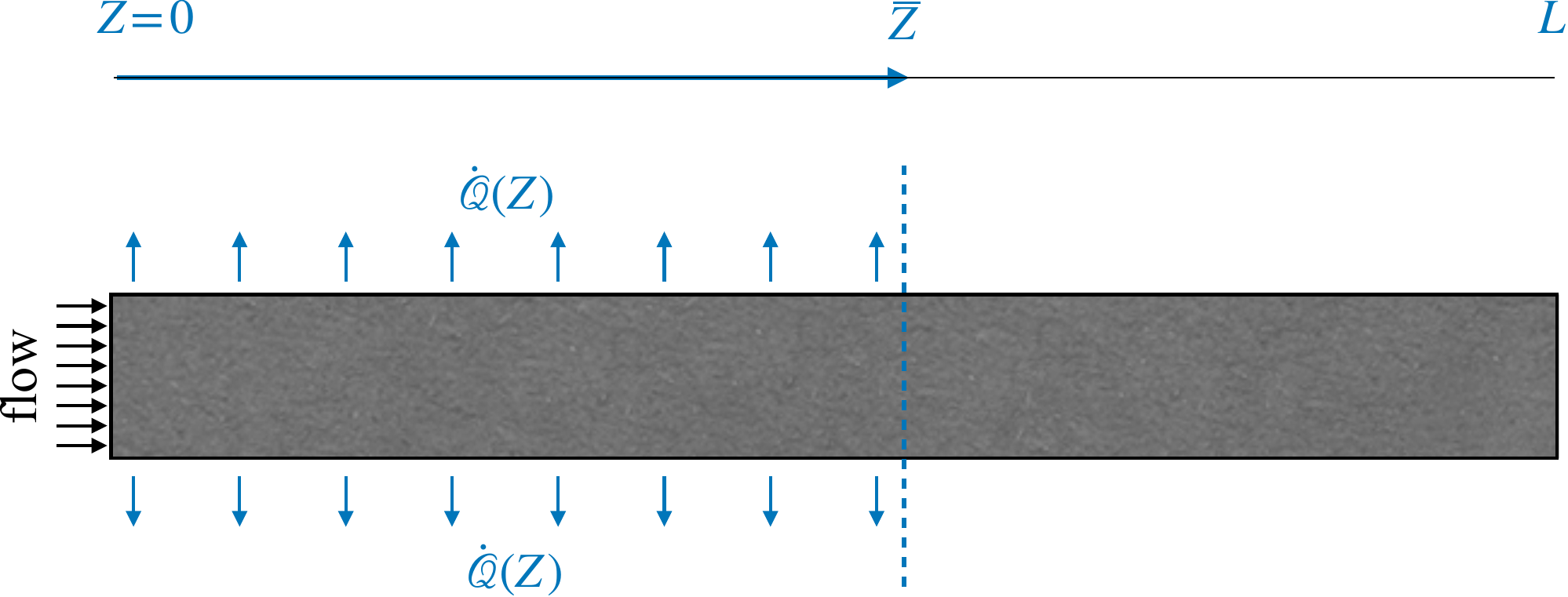}
\caption{Schematic representation of the energy balance in a generic portion $0 \leq Z \leq \overline{Z}$ of the channel.}
\label{energybalfig}
\end{figure}

Among the assumptions enforced to develop the multiscale approach discussed 
above, the isothermal condition deserves special attention.
The first issue is to determine when this assumption can be considered sensible.
If one neglects the variation of specific kinetic and potential energy, 
the steady-state energy balance 
across the channel volume between the inlet and the generic cross section at $\overline{Z}$ 
(see Fig.~\ref{energybalfig}) dictates that 
\begin{equation}
\mathcal{H}(\overline{Z})-\mathcal{H}(Z=0)=\dot{\mathcal{Q}}
\label{energybal}
\end{equation}
where $\mathcal{H}$ denotes the specific (mass-based) enthalpy 
of the gas, and where $\dot{\mathcal{Q}}(\overline{Z})$ represents the total 
heat exchanged per unit mass of fluid flowing through the channel in the given channel portion.
$0 \leq Z \leq \overline{Z}$.
Let us first consider the adiabatic case, i.e.~$\dot{\mathcal{Q}}=0$.
In this case, the temperature change downstream the channel depends on the thermal and volumetric behavior
of the gas through the  
Joule-Thomson coefficient ${\alpha}_{\rm JT}=(1/{c}_p) \, \left (T \big ({\partial} v / {\partial} T \big )_P -v \right )$
where ${c}_p$ is the constant-pressure specific heat and $v$ is the specific volume. Thus the heating/cooling effect of the gas
is determined by the heat capacity an the EOS of the fluid, which, in this context, is written in terms of large
scale variables as $f(R(Z)/M_w,P(Z),T(Z))=0$. For an ideal gas, ${\alpha}_{\rm JT}$ is 
identically vanishing at all pressures and temperatures, so that the adiabatic 
steady flow is also isothermal. Thus, in 
this case, the multiscale
approach developed above applies with no modification, provided that the dependence of the dynamic viscosity on the
pressure can be neglected.
If the latter assumption holds true, then the $Re$ number is constant,
so that Eq.~(\ref{dimnseq}) maintains its independence of the specific position $Z_h$ of the
periodic cell. 
Likewise, the local pressure drop is still given by Eq.~(\ref{pgradeq1}).
If ${\alpha}_{\rm JT}$ is different from zero, temperature changes may arise; however, even in this case,
the enthalpy conservation equation enforced between the entrance ($Z=0$) 
and a generic cross-section of $\overline{Z}$,  together with the EOS of the fluid provide a system
of two independent equations that allows to single out the dependence of the large-scale density on the large-scale
pressure $R(Z)=F(P(Z))$. Depending on the specific form of the function $F$, different
qualitative dependences $P(Z)$ can be found. 
If one were to add the effect of changes in specific kinetic energy, Eq.\ref{energybal} would become
\begin{equation}
	\mathcal{H}(\overline{Z})+\frac{R(\overline{Z}) \, V^2(\overline{Z})}{2}-\mathcal{H}(0)-\frac{R(0) \, V^2(0)}{2}=\dot{\mathcal{Q}}
	\label{energybal_kin}
\end{equation}
where the average velocity values at the entrance of the channel and in the generic cross-section $\overline{Z}$ may be related through the continuity equation to the fluid density, which can be again related to local temperature and pressure values through the fluid's EOS.

In general, this argument can also be extended to the non-adiabatic case,
provided that the heat-per-unit mass appearing at the right hand side of the enthalpy balance
can be estimated along the channel axis.
By these observations, it becomes clear that the cornerstone of the multiscale  approach is the constancy of the Reynolds 
number along the axial coordinate, which, by mass conservation, is one-to-one with the 
assumption of constant dynamic viscosity of the fluid with
temperature and pressure, an approximation that is not unreasonable in many practical cases.
To give a quantitative example, nitrogen dynamic viscosity varies by less that 40\% for $300 \leq T \leq 500$ K and $1 \leq P \leq 20$ atm.
\subsection{Extension to random media}
Another important issue to be addressed  is the possible extension of the approach to channels embedding granular packings or
random monolithic structures, which, in many applications, constitute the simplest means to obtain at one time 
high values of the surface-to-volume ratio
and of the heat and mass transfer coefficients (the latter effect being the
result of convection-enhanced transversal transport). 

In this case, the concept of periodic cell must 
be substituted with that of a statistically representative cell, that is, a portion of
length ${\Delta}{\ell}$ of the channel that is long enough as to make the values 
of the local pressure drop across the cell 
independent  of the position of the cell along the channel axis. When the overall length of the channel
becomes orders of magnitude larger than ${\Delta}{\ell}$, it is sensible to assume that the multiscale approach should be capable of capturing the
large-scale variations of pressure, density and velocity. 
Clearly, given that in this case the length ${\Delta}{\ell}$ is not an unconstrained
design  parameter but rather results from the  
features of the disordered medium (be it granular of monolithic), 
the actual value of the separation of scale parameter must be assessed on a case by case
basis. 
\section{Conclusions}
Gas flows through periodic channel geometries exhibiting large pressure drops represent a computational challenge in that,
unlike the incompressible case, 
the flow structure cannot be directly predicted by solving a single cell problem defined on the minimal periodic module of the structure.
This makes the numerical solution of the flow troublesome, or even unattainable 
when complex 3d structures and relatively high values of the
Reynolds number are entailed.
We show that a  physically sound approximation to the flow, which becomes more and more accurate as the separation of scale increases, can be
obtained by a multiscale approach, where the gas velocity is split into a large-scale piecewise-constant factor, times a strictly periodic
incompressible component. This allows to obtain a numerical prediction of the flow structure into two steps: -(i) solving an incompressible Navier-Stokes problem
on the minimal modulus of periodicity of the structure, and -(ii) obtaining the large-scale profiles of the dependent variables by analytical integration
(quadratures) of the local pressure drop across the unit cell. The method can be generalized with minimal modifications to include
the non-isothermal case, provided that the dynamic viscosity of the gas can be considered constant in the range
of temperatures and pressures interested. Different qualitative large-scale pressure/density/velocity profiles can be
obtained depending on the structure of the equation of state of the compressible fluid and/or of the heat transfer conditions along the channel axis.
The approach is validated by comparing the prediction based on the multiscale approach with the full solution
of the compressible isothermal flow of an ideal gas through different axial-symmetric periodic channel geometries. The comparison shows 
that when the ratio ${\Delta}{\ell}/L$ yielding the length of the periodic module to the overall channel length falls below $10^{-3}$, the large
scale pressure is predicted with a relative error of the order of few percent. More importantly, the local flow structure
(at the scale of the periodic cell) is also accurately predicted, thus allowing the screening of a wide number of possible
different cell geometries for tailored applications. The case of a cylindrical capillary embedding and helicoidal baffle
is discussed in some detail as an example of 3d geometry where the enhancement of mass and heat transport coefficients can be
expected due to the presence of transversal vortices, which become more and more intense with respect
to the axial velocity as the Reynolds number
increases. 
The natural extension of the approach presented in this article should be oriented towards random media, 
where the concept of periodic module
should be substituted by that of statistically representative cell, that is, 
of a portion of the channel that is long enough as to make
the local dimensionless pressure drop 
independent of its position along the channel axis. 
 
\appendix
\section{Decomposition of the solution}
\label{appA}
\begin{figure}[H]
\includegraphics[scale=0.6]{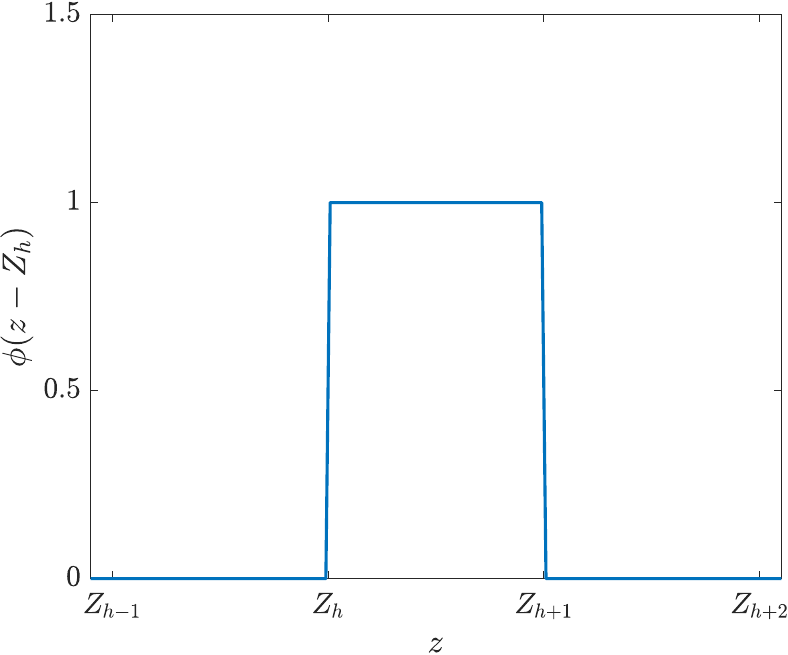}
\centering
\caption{Characteristic function of the generic $h$-th cell.}
\label{figA1}
\end{figure}

\begin{figure}[H]
\includegraphics[scale=0.4]{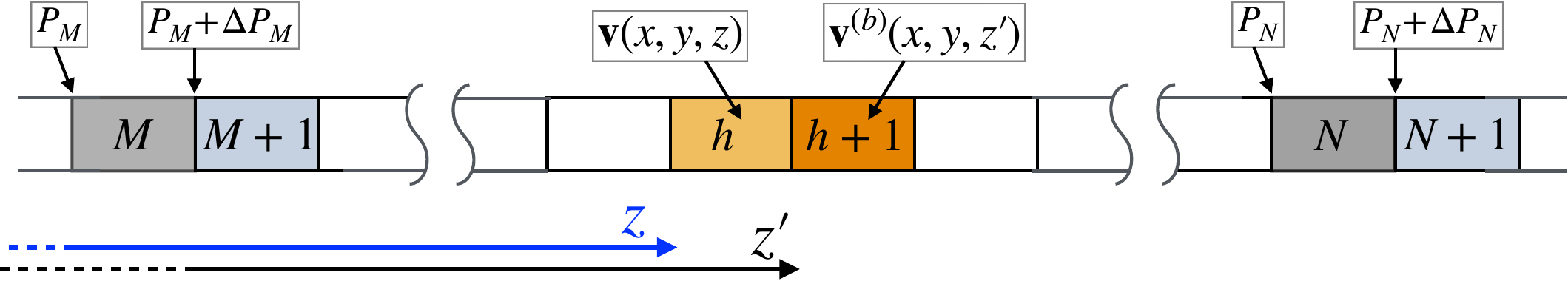}
\centering
\caption{Schematic representation of the reference frames $z$ and $z^{\prime}$
used
in Eqs.(\ref{eqA5}) through (\ref{eqA9}).}
\label{figA2}
\end{figure}

In order to investigate the conditions under which $ \mathbf{v}_c(x,y,\tilde{z})$ in eq. (\ref{vspliteq}) is periodic, it is convenient to
express the assumption eq. (\ref{rhospliteq}) on the density field $\rho(x,y,z)$ in terms of a characteristic function. Specifically,
we express the density field as
\begin{equation}
\rho(x,y,z)=R(z)
= \sum_h^{N} R_h \, \phi(z-Z_h) 
\label{eqA1}
\end{equation}
where $R_h=R(Z_h)$ is the density of the fluid in the $h$-th cell and $\phi(z-Z_h)$ is a distribution with 
compact support in the $h$-th cell depicted in Figure \ref{figA1}.
The characteristic function $\phi(z-Z_h)$ can be written 
as the difference of step (Heaviside) functions as 
\begin{equation}
 \phi(z-Z_h) = H(z-Z_h)-H(z-Z_{h+1})
\label{eqA2}
\end{equation}
Using the expression for the density eq. (\ref{eqA1}), the Navier-Stokes problem eq. (\ref{nseq}) 
becomes
\begin{equation}
\displaystyle
\begin{array}{l}
 R(z) \mathbf{v}(x,y,z) \cdot {\nabla}   \mathbf{v}  (x,y,z)
={\mu}{\nabla}^2 \mathbf{v}(x,y,z)+\frac{\mu}{3} \, 
{\nabla} \big ( {\nabla} \cdot \mathbf{v}(x,y,z) \big ) - {\nabla}p(x,y,z) \mbox{;} 
\\
[15pt]
\displaystyle
\nabla \cdot \mathbf{v}(x,y,z)
=- {v}_z(x,y,z) 
\frac{1}{R(z)}\frac{d R(z)}{dz}
\end{array}
\label{eqA3}
\end{equation}
where
\begin{equation}
\frac{d R(z)}{dz} =  \sum_h^{N_{\rm tot}} R_h \left( \, \delta(z-Z_h)-\delta(z-Z_{h+1}) \right)
\label{eqA4}
\end{equation}
and $ \delta(z-Z_h)$ is the Dirac delta function.
Considering the problem (\ref{eqA3}) in the $h$-th cell and considering
$h>>1$ large, the boundary conditions can be expressed as
\begin{equation}
p(x,y,M \Delta \ell)=P_{M};\quad p(x,y,N \Delta \ell)=P_{N}
\label{eqA5}
\end{equation}
with vanishing velocity at the fluid-solid interfaces,
where $P_M$ is the inlet pressure on the $M$-th cell,
 $P_N$ is the inlet pressure on the $N$-th cell,
with $1 \leq M \ll h$ and $h \ll  N \leq N_{\rm tot}-1$, $N_{\rm tot}$ being
the total number of cells over the entire channel length
(see Fig.~\ref{figA2}).
\vskip 0.6cm
\vskip 0.6cm

Without loss of generality, we can define a new Navier-Stokes solution
on the $(h+1)$-th cell and depending on the shifted coordinate $z'=z-\Delta \ell$ such that
\begin{equation}
\Big(\mathbf{v}^{(b)}(x,y,z'), p^{(b)}(x,y,z')\Big)=
\Big(\mathbf{v}(x,y,z+\Delta\ell), p(x,y,z + \Delta \ell)\Big)
\label{eqA6}
\end{equation}

Furthermore,
due to the periodicity of the geometry, the problem Eqs. (\ref{eqA3}) and (\ref{eqA5}) can be equivalently defined by enforcing the appropriate pressures on the adjacent $M+1$ and $N+1$ cells. Therefore,
the equations governing the auxiliary solution $(\mathbf{v}^{(b)}(x,y,z'), p^{(b)}(x,y,z'))$ read
\begin{equation}
\begin{array}{l}
\displaystyle
 R(z') \mathbf{v}^{(b)}(x,y,z') \cdot {\nabla}   \mathbf{v}^{(b)}(x,y,z')
={\mu}{\nabla}^2 \mathbf{v}^{(b)}(x,y,z')+\frac{\mu}{3} \, 
{\nabla} \big ( {\nabla} \cdot \mathbf{v}^{(b)}(x,y,z') \big ) - {\nabla}p^{(b)}(x,y,z') \mbox{;} 
\\
[15pt]
\displaystyle
\nabla \cdot \mathbf{v}^{(b)}(x,y,z')
=- {v}^{(b)}_z(x,y,z') 
\frac{1}{R(z')}\frac{d R(z')}{dz}
\end{array}
\label{eqA7}
\end{equation}
with boundary conditions
\begin{equation}
p^{(b)}(x,y,M \Delta \ell)=P_{M}+\Delta P_M;\quad p^{(b)}(x,y,N \Delta \ell)= P_{N}+\Delta P_N
\label{eqA8}
\end{equation}
where  $\Delta P_M= P_{M+1}-P_{M} $ and $\Delta P_N= P_{N+1}-P_{N} $.
By defining the pressure field 
\begin{equation}
\tilde{p}^{(b)}(x,y,z')={p}^{(b)}(x,y,z')- \Delta P_M
\label{eqA9}
\end{equation}
we have
\begin{equation}
\displaystyle
\begin{array}{l}
 R(z') \mathbf{v}^{(b)}(x,y,z') \cdot {\nabla}   \mathbf{v}^{(b)}(x,y,z')
={\mu}{\nabla}^2 \mathbf{v}^{(b)}(x,y,z')+\frac{\mu}{3} \, 
{\nabla} \big ( {\nabla} \cdot \mathbf{v}^{(b)}(x,y,z') \big ) - {\nabla}\tilde{p}^{(b)}(x,y,z') \mbox{;} 
\\
[15pt]
\displaystyle
\nabla \cdot \mathbf{v}^{(b)}(x,y,z')
=- {v}^{(b)}_z(x,y,z') 
\frac{1}{R(z')}\frac{d R(z')}{dz}
\end{array}
\label{eqA10}
\end{equation}
with boundary conditions
\begin{equation}
\tilde{p}^{(b)}(x,y,M \Delta \ell)=P_{M};\quad \tilde{p}^{(b)}(x,y,N \Delta \ell)= P_{N}+(P_{N+1}-P_{M+1})-(P_{N}-P_{M}) 
\label{eqA11}
\end{equation}
Since $ (P_{N+1}-P_{M+1})$ and $(P_{N}-P_{M}) $ are the pressure
drops on an equal large number of arbitrary chosen equivalent cells, 
we can consider them be equal each other. Therefore,
\begin{equation}
\tilde{p}^{(b)}(x,y,M \Delta \ell)=P_{M};\quad \tilde{p}^{(b)}(x,y,N \Delta \ell)= P_{N}
\label{eqA12}
\end{equation}
Finally the density $R(z')$ is related to the pressure of the cell $P_h$ by an Equation Of State of the fluid. At constant temperature
\begin{equation}
 R(z') = F(P_{h}+ \Delta P_h) 
\end{equation}
Introducing the isothermal compressibility of the fluid
\begin{equation}
\beta_h = \frac{1}{R(z)} \left(
\frac{\partial R(z)}{\partial P_h}
\right)_T
\end{equation}
Therefore, for small pressure drop in the cell
\begin{equation}
 R(z') = R(z)(1 + \beta_h \Delta P_h)
 \label{eqA15}
\end{equation}
That provides
\begin{equation}
\begin{array}{l}
\displaystyle
 R(z)\left(
 1+ \beta_h \Delta P_h
	\right) 
  \mathbf{v}^{(b)}(x,y,z') \cdot {\nabla}   \mathbf{v}^{(b)}(x,y,z')
=
\\
[15pt]
\displaystyle
{\mu}{\nabla}^2 \mathbf{v}^{(b)}(x,y,z')+\frac{\mu}{3} \, 
{\nabla} \big ( {\nabla} \cdot \mathbf{v}^{(b)}(x,y,z') \big ) - {\nabla}\tilde{p}^{(b)}(x,y,z') \mbox{;} 
\\
[15pt]
\displaystyle
\nabla \cdot \mathbf{v}^{(b)}(x,y,z')
=- {v}^{(b)}_z(x,y,z') 
\frac{1}{ R(z)\left(
 1+ \beta_h \Delta P_h
	\right) }\frac{d R(z)}{dz}
\end{array}
\label{eqA16}
\end{equation}
Comparing the original problem Eqs. (\ref{eqA3}) and (\ref{eqA5}) with the problem Eqs. (\ref{eqA16}) and (\ref{eqA12}) when $\beta_h \Delta P_h << 1 $, we can conclude that they are the same problem with the same boundary conditions. Using the definitions Eq. (\ref{eqA6}) and (\ref{eqA9}), we obtain
\begin{equation}
\mathbf{v}(x,y,z+\Delta\ell)=\mathbf{v}(x,y,z)
\label{eqA17}
\end{equation}
and
\begin{equation}
p(x,y,z + \Delta \ell)=p(x,y,z )- \Delta P_M
\label{eqA18}
\end{equation}
The periodicity expressed in Eqs. (\ref{eqA17}) and (\ref{eqA18})
is fulfilled only locally. In fact, the density of the $(l+h)$-th  cell
is
\begin{equation}
 R(z') \sim R(z)\left(1 + \sum_{j=h}^l \beta_j \Delta P_j \right)
\label{eqA19}
\end{equation}
which don't fulfil the periodicity for large values of $l$.
In order to assess the periodicity of the solution across the entire channel, we have to decouple
the thermodynamics effects due to the pressure drops along the channel
from the fluid dynamics effects due to the geometry of the channel.
Without loss of generality, we can define a normalized solution ($\mathbf{v}_h(x,y,\tilde{z})$, ${p}_h(x,y,\tilde{z}) $) in the $h$-th cell, with $\tilde{z} \in [0,\Delta l]$
 such that
\begin{equation}
\mathbf{v}(x,y,z) = V(Z_h)\,  {\mathbf{ v}}_h(x,y,\tilde{z}) 
\quad {\rm for}\quad z \in [Z_h,Z_{h+1}] 
\label{eqA20}
\end{equation}
and
\begin{equation}
p(x,y,z) = P(Z_h)\, {p}_h(x,y,\tilde{z})  
\quad {\rm for}\quad z \in [Z_h,Z_{h+1}] 
\label{eqA21}
\end{equation}
Using the characteristic function depicted in Figure \ref{figA1}, the velocity field in the channel can be written as
\begin{equation}
\mathbf{v}(x,y,z) =
 \sum_h^{N_{\rm tot}} V(Z_h) \,{\mathbf{ v}}_h(x,y,\tilde{z})   \, \phi(z-Z_h) 
\label{eqA22}
\end{equation}
and
\begin{equation}
{p}(x,y,z) =
 \sum_h^{N_{\rm tot}} P(Z_h) p_h(x,y,\tilde{z}) \, \phi(z-Z_h) 
\label{eqA23}
\end{equation}
Substituting Eq. (\ref{eqA22}) into the mass conservation expression of Eqs. (\ref{eqA3}), we have
\begin{equation}
\begin{array}{l}
  \displaystyle
 R(z)
\left( 
 \sum_h^{N_{\rm tot}} V(Z_h)  \, \phi(z-Z_h)  \,\nabla \cdot {\mathbf{ v}}_h(x,y,\tilde{z}) 
 +
  \sum_h^{N_{\rm tot}} V(Z_h)  \, \phi'(z-Z_h)  \, {v}_{z,h}(x,y,\tilde{z}) 
  \right)
  =
  \\
  [20pt]
  \displaystyle
   -\sum_h^{N_{\rm tot}} V(Z_h)  \, \phi(z-Z_h)  \, {v}_{z,h}(x,y,\tilde{z}) 
   \sum_h^{N_{\rm tot}} R_h \phi'(z-Z_h)
\end{array}
\label{eqA24}
\end{equation}
From Eq. (\ref{eqA24}), 
\begin{equation}
\nabla \cdot {\mathbf{ v}}_h(x,y,\tilde{z}) = 0 
\label{eqA25}
\end{equation}
 if the condition
\begin{equation}
\begin{array}{l}
 \displaystyle
 \sum_h^{N} R_h \, \phi(z-Z_h) 
   \sum_h^{N_{\rm tot}} V(Z_h)  \, \phi'(z-Z_h)  \, {v}_{z,h}(x,y,\tilde{z}) =
   \\[20pt]
   \displaystyle
      -\sum_h^{N_{\rm tot}} V(Z_h)  \, \phi(z-Z_h)  \, {v}_{z,h}(x,y,\tilde{z}) 
   \sum_h^{N_{\rm tot}} R_h \phi'(z-Z_h)
\end{array}
\label{eqA26}
\end{equation}
is satisfied.
Integrating on $z \in [Z_h, Z_{h+1}]$, the condition fulfilling Eq. (\ref{eqA25}) reads
\begin{equation}
R_h \, \left(
V(Z_h) - V(Z_{h+1})
\right) 
=
V(Z_h)\left(R_{h+1} - R_h\right)
\label{eqA27}
\end{equation}
Or equivalently
\begin{equation}
V(Z_{h+1})
=
V(Z_{h})
\left(
1- \frac{\Delta R_h}{R_h}
\right) 
\label{eqA28}
\end{equation}
with $ \Delta R_h=R_{h+1}- R_h$. Therefore, by adopting a normalization of each cell following the sequence Eq. (\ref{eqA28})  the 
velocity field $  {\mathbf{ v}}_h(x,y,\tilde{z}) $ is incompressible.
  
Using the decomposition Eq. (\ref{eqA20})  
the expression between two adjacent cells  
in Eq. (\ref{eqA17}) becomes
\begin{equation}
V(Z_{h+1}) \, \mathbf{v}_{h+1}(x,t,\tilde{z})
=
V(Z_{h})\, \mathbf{v}_{h}(x,t,\tilde{z})
\label{eqA29}
\end{equation}
 Substituting Eq. (\ref{eqA28}), we have
\begin{equation}
\mathbf{v}_{h+1}(x,t,\tilde{z})(1-\beta_h \Delta P_h)
=
 \mathbf{v}_{h}(x,t,\tilde{z})
\label{eqA30}
\end{equation}
which, since $\beta_h \Delta P_h \ll 1$, reads 
\begin{equation}
\mathbf{v}_{h+1}(x,t,\tilde{z})
=
\mathbf{v}_{h}(x,t,\tilde{z})
\label{eqA31}
\end{equation}
for any $h$ (once  $ \beta_h \Delta P_h \ll 1$  is ensured).

Next, let us check if $  \mathbf{v}_{h}(x,t,\tilde{z})$ the Navier-Stokes equations for the moment balance.
Considering the periodicity of $\mathbf{v}_{h}(x,t,\tilde{z})$, we have
\begin{equation}
\mathbf{v}_{h}(x,t,\tilde{z})
\phi'(z-Z_h) = 0
\label{eqA32}
\end{equation}
Therefore
\begin{equation}
\nabla \left(
 \sum_h^{N_{\rm tot}} V(Z_h) \,{\mathbf{ v}}_h(x,y,\tilde{z})   \, \phi(z-Z_h)
 \right) =
  \sum_h^{N_{\rm tot}}    V(Z_h)\, \phi(z-Z_h)
  \nabla 
 \,{\mathbf{ v}}_h(x,y,\tilde{z})  
\label{eqA33}
\end{equation}
Whereas, the gradient of the pressure provides
\begin{equation}
\begin{array}{l}
\displaystyle
\nabla \left(
 \sum_h^{N_{\rm tot}} P(Z_h) \,{p}_h(x,y,\tilde{z})   \, \phi(z-Z_h)
 \right) =
 \\
 [20pt]
 \displaystyle
  \sum_h^{N_{\rm tot}}    P(Z_h)\bigg( \phi(z-Z_h)
  \nabla 
 \,{p}_h(x,y,\tilde{z})+
   \phi'(z-Z_h) \,{p}_h(x,y,\tilde{z})
   \bigg)
\end{array}
\label{eqA34}
\end{equation}
Next, we set the characteristic pressure of the cell $P(Z_h)=R_h V(Z_h)^2$ and use the scaled spatial variables $\nabla_c = \Delta\ell\, \nabla
$ and $\zeta=\tilde{z}/\Delta \ell$.
Substituting Eqs. (\ref{eqA22}) and (\ref{eqA23}) into the moment balance equation of Eqs. (\ref{eqA30}), and integrating over $[Z_h,Z_{h+1}]$, the equations
associated to $\mathbf{v}_{h}(x,t,\tilde{z})$
read
\begin{equation}
\begin{array}{l}
\displaystyle
\int_{Z_h}^{Z_{h+1}} \phi(z-Z_h) \bigg(
 \mathbf{v}_h(x,y,\tilde{z}) \cdot {\nabla}_c   \mathbf{v}_h (x,y,\tilde{z})
-
\\
[20pt]
\displaystyle
\left.
\frac{1}{{\rm Re}_h}{\nabla}_c^2 \,\mathbf{v}_h(x,y,\tilde{z})-\frac{1}{3} \, 
{\nabla}_c \big ( {\nabla}_c \cdot \mathbf{v}_h(x,y,\tilde{z}) \big ) + {\nabla}_c\, p_h(x,y,\tilde{z}) 
\right) dz
=
\Delta p_h
\mbox{;} 
\\
[15pt]
\displaystyle
\nabla \cdot \mathbf{v}_h(x,y,\tilde{z})
=0
\end{array}
\label{eqA35}
\end{equation}
where $\Delta p_h = p_h(x,y,\Delta \ell)- p_h(x,y,0)$ and ${\rm Re}_h$
is the Reynolds number of the $h$-th cell, defined as
\begin{equation}
{\rm Re}_h= \frac{R(Z_h) \, V(Z_h) \Delta l}{\mu}
\label{eqA36}
\end{equation}
Eqs. (\ref{eqA35}) is the week formulation of the problem
\begin{equation}
\begin{array}{l}
\displaystyle
 \mathbf{v}_h(x,y,\tilde{z}) \cdot {\nabla}_c   \mathbf{v}_h (x,y,\tilde{z})
=
\frac{1}{{\rm Re}_h}{\nabla}_c^2 \,\mathbf{v}_h(x,y,\tilde{z})+\frac{1}{3} \, 
{\nabla}_c \big ( {\nabla}_c \cdot \mathbf{v}_h(x,y,\tilde{z}) \big ) - {\nabla}_c\, ( p_h(x,y,\tilde{z}) - \Delta p_h\,\zeta) 
\mbox{;} 
\\
[15pt]
\displaystyle
\nabla_c \cdot \mathbf{v}_h(x,y,\tilde{z})
=0
\end{array}
\label{eqA37}
\end{equation}

From the Eq. (\ref{eqA28})
\begin{equation}
R(Z_h)\,V(Z_{h})
= R(Z_h)\,
V(Z_{h +1})\,
\left(
1- \frac{\Delta R_h}{R_h}
\right)^{-1}
\label{eqA38}
\end{equation} 
Since $ \Delta R_h / R_h < 1$, we can expand the geometric series in Eq. (\ref{eqA38}), obtaining
\begin{equation}
R(Z_h)\,V(Z_{h})
= R(Z_{h+1})\,
V(Z_{h +1})\,
\left(
1 +O (\beta_h \Delta P_h)^2 
\right)
\label{eqA39}
\end{equation} 
where the relation Eq. (\ref{eqA15}) has been used.

From Eq. (\ref{eqA39}), it is easy to prove by recursion that, for each couple of cells $j$ and $k$ belonging to the channel
\begin{equation}
R(Z_j)\,V(Z_{j})
= R(Z_{k})\,
V(Z_{k})\,
\left(
1 +O (\beta {\Delta P)}^2_{\rm max} 
\right)
\label{eqA36}
\end{equation} 
where $(\beta \Delta P)_{\rm max}$ is the maximum value of $\beta_h \Delta P_h $
over all of the $h$-th cells located between the cells $j$ and $k$.

Finally, from Eq. (\ref{eqA36}), we can consider a constant Reynolds number ${\rm Re} = {\rm Re}_h$
across the entire channel and a periodic velocity field
$  \mathbf{v}_c(x,y,\tilde{z}) = \mathbf{v}_h(x,y,\tilde{z}) $ solution of the equations 
\begin{equation}
\begin{array}{l}
\displaystyle
 \mathbf{v}_c(x,y,\tilde{z}) \cdot {\nabla}_c   \mathbf{v}_c (x,y,\tilde{z})
=\frac{1}{{\rm Re}}{\nabla}_c^2 \mathbf{v}_c(x,y,\tilde{z})+\frac{1}{3} \, 
{\nabla}_c \big ( {\nabla}_c \cdot \mathbf{v}_c(x,y,\tilde{z}) \big ) - {\nabla}_c \big( p_c(x,y,\tilde{z})- \Delta p_c\, \zeta \big) \mbox{;} 
\\
[15pt]
\displaystyle
\nabla_c \cdot \mathbf{v}_c(x,y,\tilde{z})
=0
\end{array}
\label{eqA33}
\end{equation}
By which 
\begin{equation}
\mathbf{v}(x,y,z) = V(Z_h)\,  {\mathbf{ v}}_c(x,y,\tilde{z}) 
\quad {\rm for}\quad z \in [Z_h,Z_{h+1}] 
\label{eqA34}
\end{equation}

\bibliographystyle{elsarticle-num}
\bibliography{bibliography}

\begin{thebibliography}{10}
\expandafter\ifx\csname url\endcsname\relax
  \def\url#1{\texttt{#1}}\fi
\expandafter\ifx\csname urlprefix\endcsname\relax\def\urlprefix{URL }\fi
\expandafter\ifx\csname href\endcsname\relax
  \def\href#1#2{#2} \def\path#1{#1}\fi

\bibitem{li2021high}
X.~Li, Y.~Huang, Z.~Wu, H.~Gu, X.~Chen, High conversion hydrogen peroxide
  microchannel reactors: Design and two-phase flow instability investigation,
  Chemical Engineering Journal 422 (2021) 130080.

\bibitem{li2024efficient}
C.~Li, H.~Zhang, W.~Liu, L.~Sheng, M.-J. Cheng, B.~Xu, G.~Luo, Q.~Lu, Efficient
  conversion of propane in a microchannel reactor at ambient conditions, Nature
  communications 15~(1) (2024) 884.

\bibitem{feng2022residence}
H.~Feng, R.~Chen, Residence time characteristics of taylor reacting flow in a
  microchannel reactor during long-term operation, ACS Sustainable Chemistry \&
  Engineering 10~(13) (2022) 4105--4113.

\bibitem{chen2024hydrogen}
J.~Chen, Hydrogen production in protruded millisecond microchannel reactors by
  catalytically reforming methanol, International Journal of Hydrogen Energy 67
  (2024) 225--239.

\bibitem{bucak2022heat}
H.~Bucak, F.~Yilmaz, Heat transfer augmentation using periodically spherical
  dimple-protrusion patterned walls of twisted tape, International Journal of
  Thermal Sciences 171 (2022) 107211.

\bibitem{rahman2024assessment}
M.~A. Rahman, S.~M. Hasnain, R.~Zairov, Assessment of improving heat exchanger
  thermal performance through implementation of swirling flow technology,
  International Journal of Thermofluids (2024) 100689.

\bibitem{biagioni2022taming}
V.~Biagioni, C.~Venditti, A.~Adrover, M.~Giona, S.~Cerbelli, Taming taylor-aris
  dispersion through chaotic advection, Journal of Chromatography A 1673 (2022)
  463110.

\bibitem{Farkya2025}
A.~Farkya, A.~S. Rana, Modeling of rarefied gas flows in streamwise periodic
  channels: Application of coupled constitutive relations and the method of
  fundamental solutions, Engineering Analysis with Boundary Elements 172
  (2025).

\bibitem{Kosyanchuk202290}
V.~Kosyanchuk, V.~Pozhalostin, Non-stationary rarefied gas flow in a plane
  channel with a series of oscillating barriers, European Journal of Mechanics,
  B/Fluids 92 (2022) 90 – 99.

\bibitem{Zhu2017}
L.~Zhu, Z.~Guo, Numerical study of nonequilibrium gas flow in a microchannel
  with a ratchet surface, Physical Review E 95~(2) (2017).

\bibitem{Patronis2014532}
A.~Patronis, D.~A. Lockerby, Multiscale simulation of non-isothermal
  microchannel gas flows, Journal of Computational Physics 270 (2014) 532 –
  543.

\bibitem{prud1986laminar}
R.~K. Prud'Homme, T.~W. Chapman, J.~R. Bowen, Laminar compressible flow in a
  tube, Applied scientific research 43 (1986) 67--74.

\bibitem{venerus2006laminar}
D.~C. Venerus, Laminar capillary flow of compressible viscous fluids, Journal
  of Fluid Mechanics 555 (2006) 59--80.

\bibitem{venerus2010compressible}
D.~Venerus, D.~Bugajsky, Compressible laminar flow in a channel, Physics of
  Fluids 22~(4) (2010).

\bibitem{celata2007experimental}
G.~Celata, M.~Cumo, S.~McPhail, L.~Tesfagabir, G.~Zummo, Experimental study on
  compressible flow in microtubes, International Journal of Heat and Fluid Flow
  28~(1) (2007) 28--36.

\bibitem{celata2009friction}
G.~Celata, M.~Lorenzini, G.~Morini, G.~Zummo, Friction factor in micropipe gas
  flow under laminar, transition and turbulent flow regime, International
  Journal of heat and fluid flow 30~(5) (2009) 814--822.

\bibitem{guo2008numerical}
X.~Guo, C.~Huang, A.~Alexeenko, J.~Sullivan, Numerical and experimental study
  of gas flows in 2d and 3d microchannels, Journal of Micromechanics and
  Microengineering 18~(2) (2008) 025034.

\bibitem{bejhed2006numerical}
J.~Bejhed, H.~Nguyen, P.~{\AA}strand, A.~Eriksson, J.~K{\"o}hler, Numerical
  modeling and verification of gas flow through a network of crossed narrow
  v-grooves, Journal of Micromechanics and Microengineering 16~(10) (2006).

\bibitem{novopashin2016laminar}
S.~Novopashin, P.~Skovorodko, G.~Sukhinin, Laminar-turbulent transition in
  hagen--poiseuille flow of a real gas, Journal of Turbulence 17~(9) (2016)
  870--877.

\bibitem{vocale2022numerical}
P.~Vocale, D.~Rehman, G.~Morini, Numerical investigation of compressibility
  effects on friction factor in rectangular microchannels, International
  Journal of Thermal Sciences 172 (2022) 107373.

\bibitem{su2025experimental}
X.~Su, Y.~Zhang, Y.~Rao, K.~Yeranee, X.~Wang, Experimental and numerical study
  on flow and heat transfer characteristics of additively manufactured triply
  periodic minimal surface (tpms) heat exchangers for micro gas turbine,
  Aerospace 12~(5) (2025) 416.

\bibitem{cheng2021investigations}
Z.~Cheng, X.~Li, R.~Xu, P.~Jiang, Investigations on porous media customized by
  triply periodic minimal surface: Heat transfer correlations and strength
  performance, International Communications in Heat and Mass Transfer 129
  (2021) 105713.

\bibitem{jespers2017chip}
S.~Jespers, S.~Schlautmann, H.~Gardeniers, W.~De~Malsche, F.~Lynen, G.~Desmet,
  Chip-based multicapillary column with maximal interconnectivity to combine
  maximum efficiency and maximum loadability, Analytical chemistry 89~(21)
  (2017) 11605--11613.

\bibitem{bragin2024flow}
D.~Bragin, I.~Karpilov, D.~Pashchenko, Flow dynamics through cellular material
  based on a structure with triply periodic minimal surface, Chemical
  Engineering Science 298 (2024) 120291.

\bibitem{meziani2022evaluation}
A.~Meziani, S.~Verloy, O.~Ferroukhi, S.~Roca, A.~Curat, S.~Tisse,
  V.~Peulon-Agasse, H.~Gardeniers, G.~Desmet, P.~Cardinael, Evaluation of gas
  chromatography columns with radially elongated pillars as second-dimension
  columns in comprehensive two-dimensional gas chromatography, Analytical
  chemistry 94~(41) (2022) 14126--14134.

\bibitem{liu2021behavior}
H.~Liu, J.~Randon, Behavior of micro pillar array column in high pressure gas
  chromatography, Journal of Chromatography A 1656 (2021) 462551.

\bibitem{hermann2014geometric}
H.~Hermann, A.~Elsner, Geometric models for isotropic random porous media: a
  review, Advances in Materials Science and Engineering 2014~(1) (2014) 562874.

\bibitem{thabet2018development}
A.~Thabet, A.~G. Straatman, The development and numerical modelling of a
  representative elemental volume for packed sand, Chemical Engineering Science
  187 (2018) 117--126.

\bibitem{lester2013chaotic}
D.~R. Lester, G.~Metcalfe, M.~G. Trefry, Is chaotic advection inherent to
  porous media flow?, Physical review letters 111~(17) (2013) 174101.

\bibitem{brenner1993macrotransport}
H.~Brenner, D.~A. Edwards, Macrotransport processes, edited by butterworth
  (1993).

\bibitem{poumaere2022residence}
N.~Pouma{\"e}re, B.~Pier, F.~Raynal, Residence time distributions for in-line
  chaotic mixers, Physical Review E 106~(1) (2022) 015107.

\bibitem{raynal2014distribution}
F.~Raynal, P.~Carriere, The distribution of “time of flight” in 3d
  stationary chaotic advection, arXiv preprint arXiv:1412.5295 (2014).

\bibitem{lester2018simultaneous}
D.~R. Lester, B.~Kuan, G.~Metcalfe, Simultaneous optimisation of residence
  time, heat and mass transfer in laminar duct flows, Chemical Engineering
  Science 191 (2018) 511--524.

\bibitem{gorodetskyi2014analysis}
O.~Gorodetskyi, M.~F. Speetjens, P.~D. Anderson, M.~Giona, Analysis of the
  advection--diffusion mixing by the mapping method formalism in 3d open-flow
  devices, AIChE Journal 60~(1) (2014) 387--407.

\bibitem{habchi2013chaotic}
C.~Habchi, J.-L. Harion, S.~Russeil, D.~Bougeard, F.~Hachem, A.~Elmarakbi,
  Chaotic mixing by longitudinal vorticity, Chemical Engineering Science 104
  (2013) 439--450.

\bibitem{jilisen2013three}
R.~Jilisen, P.~Bloemen, M.~Speetjens, Three-dimensional flow measurements in a
  static mixer, AIChE journal 59~(5) (2013) 1746--1761.

\bibitem{gelin2021reducing}
P.~Gelin, D.~Maes, W.~De~Malsche, Reducing taylor-aris dispersion by exploiting
  lateral convection associated with acoustic streaming, Chemical Engineering
  Journal 417 (2021) 128031.

\bibitem{de2012realization}
W.~De~Malsche, J.~Op~De~Beeck, S.~De~Bruyne, H.~Gardeniers, G.~Desmet,
  Realization of 1$\times$ 106 theoretical plates in liquid chromatography
  using very long pillar array columns, Analytical chemistry 84~(3) (2012)
  1214--1219.

\bibitem{majda1999simplified}
A.~J. Majda, P.~R. Kramer, Simplified models for turbulent diffusion: theory,
  numerical modelling, and physical phenomena, Physics reports 314~(4-5) (1999)
  237--574.

\bibitem{landau1987fluid}
L.~D. Landau, E.~M. Lifshitz, Fluid Mechanics: Volume 6, Vol.~6, Elsevier,
  1987.

\end{thebibliography}
\end{document}